\documentclass[12pt,preprint]{aastex}

\usepackage{amsmath}

\newcommand{\myemail}{quanz@mpia.de}

\slugcomment{To appear in ApJ}

\shorttitle{Dust and ices in MIR spectra of FUors}
\shortauthors{Quanz et al.}

\begin{document}


\title{Evolution of dust and ice features around FU Orionis objects\altaffilmark{1}}


\author{S. P. Quanz, Th. Henning, J. Bouwman, R. van Boekel, A. Juh\'asz, H. Linz,}
\affil{Max Planck Institute for Astronomy, K\"onigstuhl 17, Heidelberg,
    Germany}
\email{\myemail}
\author{K.~M. Pontoppidan}
\affil{California Institute of Technology, Division of Geological and Planetary Sciences, Pasadena, CA; Hubble Fellow}
\and
\author{F. Lahuis}
\affil{Leiden Observatory, P.O. Box 9513, 2300 RA Leiden, Netherlands}
\affil{SRON Netherlands Institute for Space Research, Groningen, Netherlands}

\altaffiltext{1}{Based on observations with ISO, an ESA project with instruments funded by ESA Member States (especially the PI countries: France, Germany, the Netherlands and the United Kingdom) and with the participation of ISAS and NASA. This work is based in part on observations made with the {\sc Spitzer Space Telescope}, which is operated by the Jet Propulsion Laboratory, California Institute of Technology under a contract with NASA.}







\begin{abstract}
We present spectroscopy data  
for a sample of 14 FUors and 2 TTauri stars observed with the 
{\sc Spitzer Space Telescope} or with the {\sc Infrared Space Observatory (ISO)}.
Based on the appearance of the 10\,$\mu$m silicate feature we define 2 categories of FUors.
Objects showing the silicate feature in absorption (Category 1) are still embedded in a dusty and icy
envelope. The shape of the 10\,$\mu$m silicate absorption bands is compared to typical dust compositions  
of the interstellar medium and found to be in general agreement. Only one object (RNO 1B) appears to be too
rich in amorphous pyroxene dust, 
but a superposed emission feature can explain the observed shape. We
derive optical depths and extinction values from the silicate band and additional ice bands at 
6.0, 6.8 and 15.2\,$\mu$m. In particular the analysis of the CO$_2$ ice band at 15.2\,$\mu$m allows us to 
search for evidence for ice processing and constrains whether the absorbing material 
is physically linked to the central object or in the foreground. For objects showing the 
silicate feature in emission (Category 2), we argue that the emission comes from the surface layer 
of accretion disks. Analyzing the dust composition reveals that significant grain growth has already 
taken place within the accretion disks, but no clear indications for crystallization are present.
We discuss how these observational results can be explained in the picture of a young, and highly active 
accretion disk. Finally, a framework is proposed as to how the two categories of FUors can be understood 
in a general paradigm of the evolution of young, low-mass stars. Only one object (Parsamian 21) shows
PAH emission features. Their shapes, however, are 
often seen toward evolved stars and we question the object's status as a FUor and discuss
other possible classifications. Two spectra (RNO 1B and L1551 IRS 5) show [Fe II] emission lines
which are attributed to hot and dense material located close to the root of an outflow (L1551 IRS 5) or 
to shocked material directly within an outflow (RNO 1B).
\end{abstract}





\section{Introduction}
By means of mid-infrared (MIR) spectroscopy gaseous and solid state features have been observed and
analyzed in a variety of astronomical environments. While ground-based observations are restricted
to certain atmospheric windows (e.g., the N-band around 10\,$\mu$m), spectrographs onboard  
space-borne telescopes, such as {\sc ISO} and {\sc Spitzer}, 
enable us to study a broad wavelength range from the 
near-infrared (NIR) to well beyond 30\,$\mu$m. Those instruments fostered and revolutionized 
our understanding in numerous astronomical fields of research. 
In particular the star-formation community benefitted from studies based on data from the space telescopes.
The composition of dust grains and their evolution (e.g., grain growth and crystallization) in protoplanetary 
disks was analyzed to great extend in young, intermediate mass Herbig Ae/Be stars (HAeBes) \citep[e.g.,][]{bouwman2001,meeus2001,acke2004,vanboekel2005}, but also in the young, less massive TTauri 
stars \citep[e.g.,][]{forrest2004,kessler2006,sargent2006}. In addition, the ice and dust 
features of younger and more deeply embedded objects were studied \citep[e.g.,][]{watson2004} and 
the ice inventory of molecular clouds was investigated \citep[e.g.,][]{knez2005,bergin2005}. 

One special sub-group of young low-mass objects are FU Orionis objects (FUors), named after the prototype
FU Ori. For most objects of this group evidence was found for a tremendous outburst in optical or NIR light
over short timescales (months to years) followed by a decline in luminosity typically over several decades. 
Other objects were included to this group as they shared peculiar spectral features, e.g, CO bandhead absorption 
in the NIR and a changing spectral type with wavelength. Most observational data can be well explained with 
highly active accretion disks surrounding these objects, possibly fed with fresh material from
a remnant envelope \citep[for a review, see,][]{hartmann1996}. Observations in the MIR are thus ideally suited 
to probe the dusty component of the circumstellar material of these objects, either contained
in the accretion disks or, additionally, in the envelopes. Recent findings from NIR and MIR 
interferometers support the presence of accretion disks several tens of AU in size \citep{quanz2006,malbet2005,millan-gabet2006}. 
However, while the SED and the NIR and MIR visibilities of FU Ori itself can be sufficiently explained 
by a simple accretion disk model \citep{quanz2006}, \citet{millan-gabet2006} found that accretion disks alone can not reproduce the SED and observed low K-band visibilities of V1057 Cyg, V1515 Cyg and Z CMa-SE simultaneously. 
They concluded that additional uncorrelated flux is required possibly arising due to scattering by large dusty envelopes.  

While \citet{lorenzetti2000} presented far-infrared spectroscopy data for 6 FUors observed with  
ISO-LWS, a dedicated MIR study of a larger sample of FUors is still missing.
\citet{larsson2000} showed the ISO-SWS spectra for 6 FUors for comparison, 
but no analysis was carried out. The ISO data for Z CMa were presented in publications related 
to Herbig Ae/Be stars by \citet{acke2004} and \citet{elia2004},
as Z CMa presumably is a binary system consisting of a Herbig star and a FUor. \citet{white2000} used the ISO observations of L1551 IRS 5 as input for a radiative transfer model of this source. 
\citet{hanner1998} discussed ground-based 8-13$\mu$m spectra for four FUors (FU Ori, V1515 Cyg, V1057 Cyg, V1735 Cyg) and fitted a simple dust model to the data to check whether silicate particles from the interstellar medium (ISM) can reproduce the observed features. 
\citet{schuetz2005} published additional ground-based data for another four objects in the same wavelength regime.  As three objects (Z CMa, V346 Nor, V883 Ori) showed the 10$\mu$m feature in absorption they inferred the optical depth from fitting an ISM dust model to the spectra. The fourth object (Bran 76, alias BBW 76) was not 
analyzed in greater detail. \citet{polomski2005} presented data on RNO 1B, Z CMa, and Parsamian 21 and derived
dust temperatures and optical depths. 

To our knowledge, the first MIR spectra of FUors observed with {\sc Spitzer} were presented in 
\citet{green2006}. The main focus of this publication was an accurate SED modeling of FU Ori, V1515 Cyg, and V1057 Cyg. Spectral solid state features were not analyzed in greater detail. 
The spectrum of V346 Nor was presented for comparison.   
In \citet{quanz2006} the first detailed dust composition modeling for the FU Ori spectrum was presented
and evidence for grain growth in the accretion disk was found. The {\sc Spitzer} 
spectra of RNO 1B/1C were shown in \citet{quanz2007rno}. However, the dust composition was not yet analyzed in detail.
 
In this paper we compile MIR spectra for 14 FU Orionis objects 
observed with ISO and/or {\sc Spitzer}. 
As up to now only $\sim$\,20 FUors or FUor candidates are known, 
this is the largest sample of these objects analyzed so far in a single MIR study.
Part of the data have not been published before. 
For objects where the spectra show a sufficient signal-to-noise ratio, 
the dust and ice composition of the circumstellar material is investigated. 


\section{Observations and Data Reduction}
The mid-infrared spectra we present in this paper are compiled from the archives of the ISO\footnote{The ISO archive can be accessed via http://www.iso.vilspa.esa.es/ida/} and the 
{\sc Spitzer}\footnote{http://ssc.spitzer.caltech.edu/archanaly/archive.html} satellite. 
The Short Wavelength Spectrograph (SWS) onboard {\sc ISO} consisted of two nearly independent grating
spectrometers with a spectral resolution between 1000 and 2500 (depending on the band and order) and covering a wavelength range from 2.4\,-\,12.0\,$\mu$m and from 12.0\,-\,45.0\,$\mu$m, respectively. 
While the field-of-view (FOV) for the shorter wavelength regime was 14$''\times 20''$, the FOV for the 
longer wavelength range was 14$''\times 27''$, with the exception of the wavelength range between 28.9\,-\,45.0\,$\mu$m which had a FOV of 20$''\times 33''$.
With {\sc ISO/SWS} seven objects classified as FUors were observed between April 1996 and October 1997. For one object (OO Ser) data were taken at five different epochs documenting a decay in luminosity over a few months. Table~\ref{isojournal} summarizes the {\sc ISO}-observations with object names, coordinates, {\sc ISO-SWS} observing mode and scan speed, integration time on target, possible pointing offsets (see below), and the date of the observation. 
For the data reduction Highly Processed Data Products (HPDP) or SWS Auto Analysis Results (AAR) were downloaded from the {\sc ISO} archive for speed 1 and 2 or speed 3 and 4 observations, respectively. With the OSIA software package (version 4.0)\footnote{http://sws.ster.kuleuven.ac.be/osia/} the following reduction steps were carried out:  
For each object the spectra from the {\sc ISO-SWS} up- and down-scan were flat-fielded and rebinned. 
After sigma clipping, the speed 3 and speed 4 spectra were de-fringed. This procedure was not required for the speed 1 and speed 2 data as the HPDP are already de-fringed. Finally, the spectra from the up- and down-scan were combined and rebinned to a spectral resolution of 100. In case the resulting spectrum showed signs of a pointing offset (e.g., aperture jumps) a correction based on the measured beam profiles along the different axes was applied to the raw data and the data reduction was repeated. The applied offsets are listed in Table~\ref{isojournal}.

The {\sc Spitzer} observations are summarized in Table~\ref{spitzerjournal}. 
The IRS onboard {\sc Spitzer} offers a short wavelength, low resolution module (SL) covering the wavelength
range between 5.2\,-\,14.5\,$\mu$m, and a short wavelength, high resolution module (SH) going from 9.9\,-\,19.6\,$\mu$m. The corresponding slit sizes are $\approx$\,3.6$''\times 136.0''$ 
(including both SL orders and an additional bonus segment connecting both orders) and 4.7$''\times 11.3''$, respectively. 
For the longer wavelength part a long wavelength, low resolution module (LL) ranging from 
14.0\,-\,38.0\,$\mu$m and a long wavelength, high resolution model (LH) covering the regime from 18.7\,-\,37.2\,$\mu$m are available. The slits sizes are $\approx$\,10.5$''\times 360.0''$ 
(including both LL orders and an additional bonus segment connecting both orders) and 11.1$''\times 22.3''$, respectively. Both low resolution modules (SL and LL) offer a spectral resolution between 64\,-\,128 (depending on the wavelength) while the high resolution modules both have $\sim$\,600.
All objects listed in Table~\ref{spitzerjournal} were observed with the full wavelength coverage, either with 
a combination of SL+LL or with SL+SH+LH.
Two objects (HL Tau and XZ Tau) are not classified as FUors but were part of a small {\sc Spitzer}/IRS map including L1551 IRS 5, and data for all three objects could be downloaded simultaneously. Interestingly, XZ Tau is a binary system that recently was found to show EXor-type variations \citep{coffey2004}, i.e., another type of short term eruptions of young stars. Thus, a comparison to the FUor data presented here is reasonable. In addition, the data of HL Tau enable us to compare the FUor spectra to that of a well-studied young star with a remnant envelope and a highly inclined accretion disk seen almost edge-on \citep{close1997}. Part of the {\sc Spitzer} spectrum of HL Tau was already published in \citet{bergin2005}.
The object V1647 Ori was observed three times
within a period of roughly 5 months to monitor its brightness as it underwent an eruption beginning of 2004.
However, to our knowledge, thus far no spectrum was published.

The data reduction process of the {\sc Spitzer} data is described in detail 
in \citet{quanz2007rno}. However, we should mention that the error bar for each individual spectral point
represents the formal standard deviation from the mean flux of at least two independent measurements 
(two telescope nod position and possibly several observation cycles). Also taken into account is the formal
error of the spectral response function. For details on the method we refer the reader also to \citet{bouwman2006}.
We estimate a relative flux calibration error across a spectral order of $\approx 5$\% and an absolute calibration error between orders or modules of $\approx 10$\%.
In particular, for the objects RNO 1B and RNO 1C apparent flux density offsets between the SL and the SH part of the spectra, as well as between the short and long wavelengths part of the high-resolution spectra, are 
already discussed in \citet{quanz2007rno}. 
For the other objects presented here, the discrepancies in the flux densities between the SL and 
the SH part of the spectrum were $<$\,10\% and we matched the longer wavelength part to the shorter regime by multiplying a scalar factor. Only Bran 76 (also known as BBW 76) showed a larger offset 
of $\approx$\,15\% as already mentioned by \cite{green2006}. 
For Parsamian 21 (HBC 687) we do not show the SH spectrum between 
14 and 20\,$\mu$m as the slit of the spectrograph was not centered on the source and significant flux loss
occurred for which we could not correct. An additional correction between the SH and the LH module of the spectrum was
required for XZ Tau, HL Tau and L1551 IRS 5 where the flux densities of the LH part had to be scaled down 
by 10\,-\,15\%. This offset can be explained by the larger aperture of this module which possibly probed 
additional large scale emission from the surroundings of these objects. 


\section{Results}
\subsection{General overview}
\begin{figure}
\centering
\epsscale{1.}
    \plotone{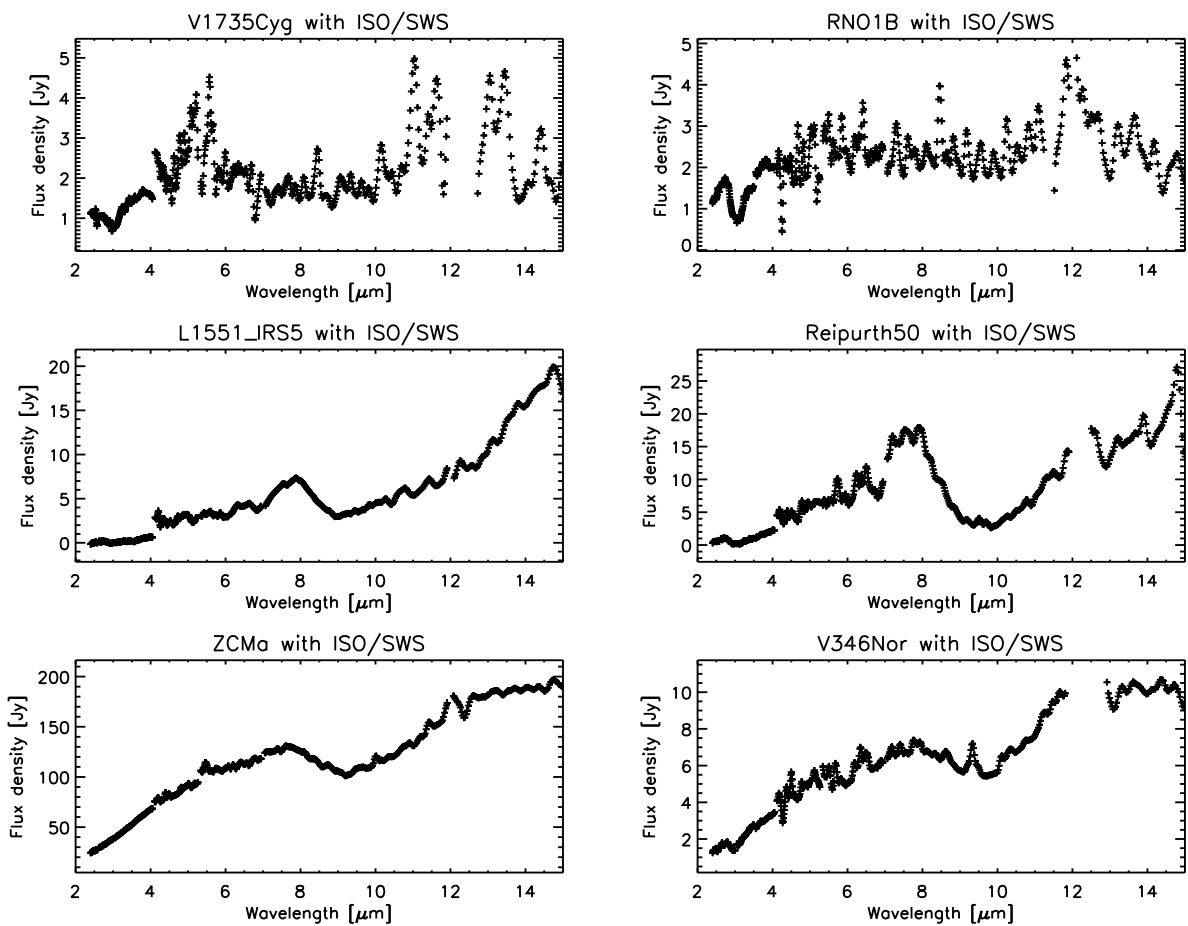}
   \caption{ISO-SWS spectra in the 2-15\,$\mu$m range for 6 of our targets. The two 
   plots in the top panel (V1735 Cyg and RNO 1B) suffer from a poor signal-to-noise 
ratio resulting from a combination of relatively low flux levels and a short integration times (Table~\ref{isojournal}). All sub-structures in the spectra are not believed to be real. }
 \label{iso_short1}
\end{figure}

\begin{figure}
\centering
\epsscale{1.}
    \plotone{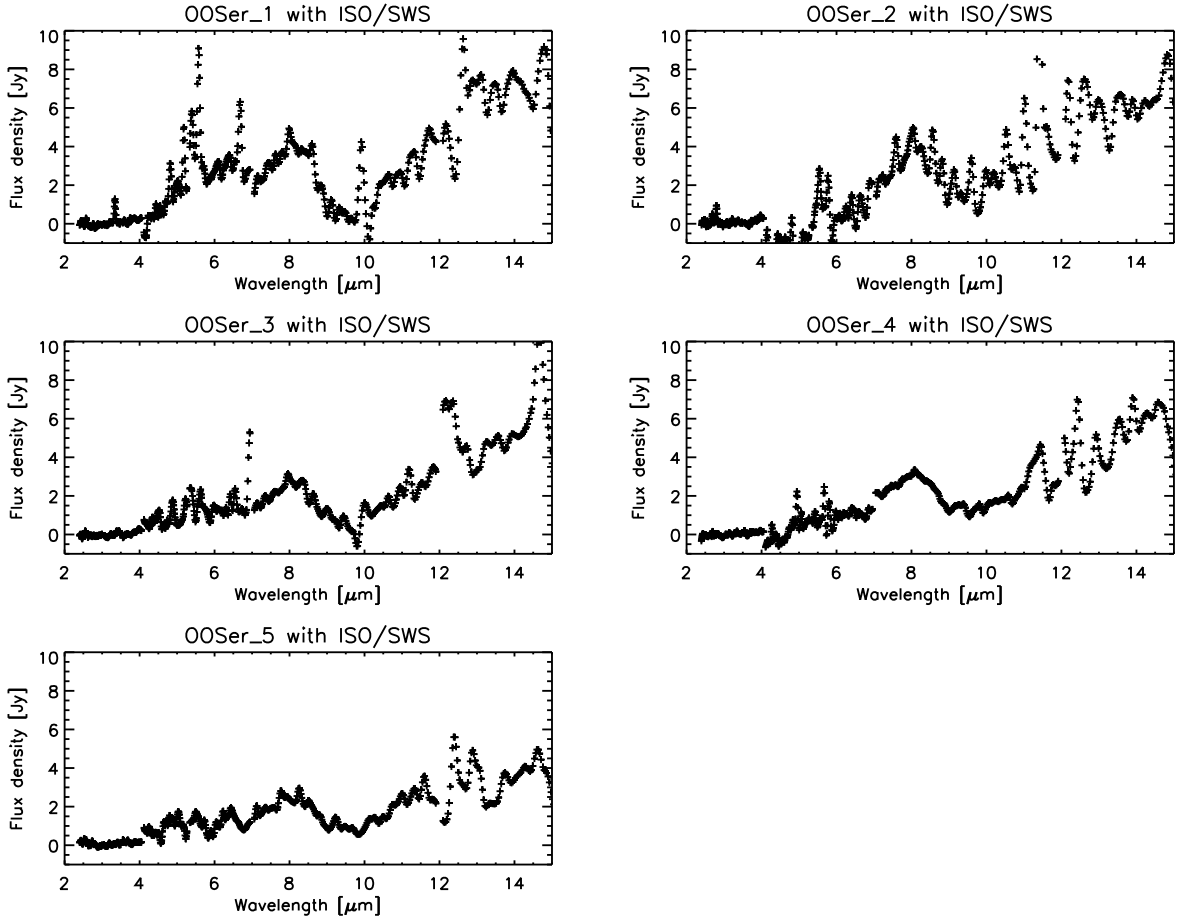}
   \caption{Same as Figure~\ref{iso_short1} but for the object OO Ser which was observed at 5 
   different epochs. As shown in Table~\ref{isojournal} the integration time of the 
   observations were increased during the 5 epochs. However, as the object faded significantly over this period,
   the signal-to-noise ratio could not be improved and it is difficult to identify spectroscopic features in the data
   apart from the silicate absorption profile seen in all data sets.}
 \label{iso_short2}
\end{figure}

\begin{figure}
\centering
\epsscale{1.}
    \plotone{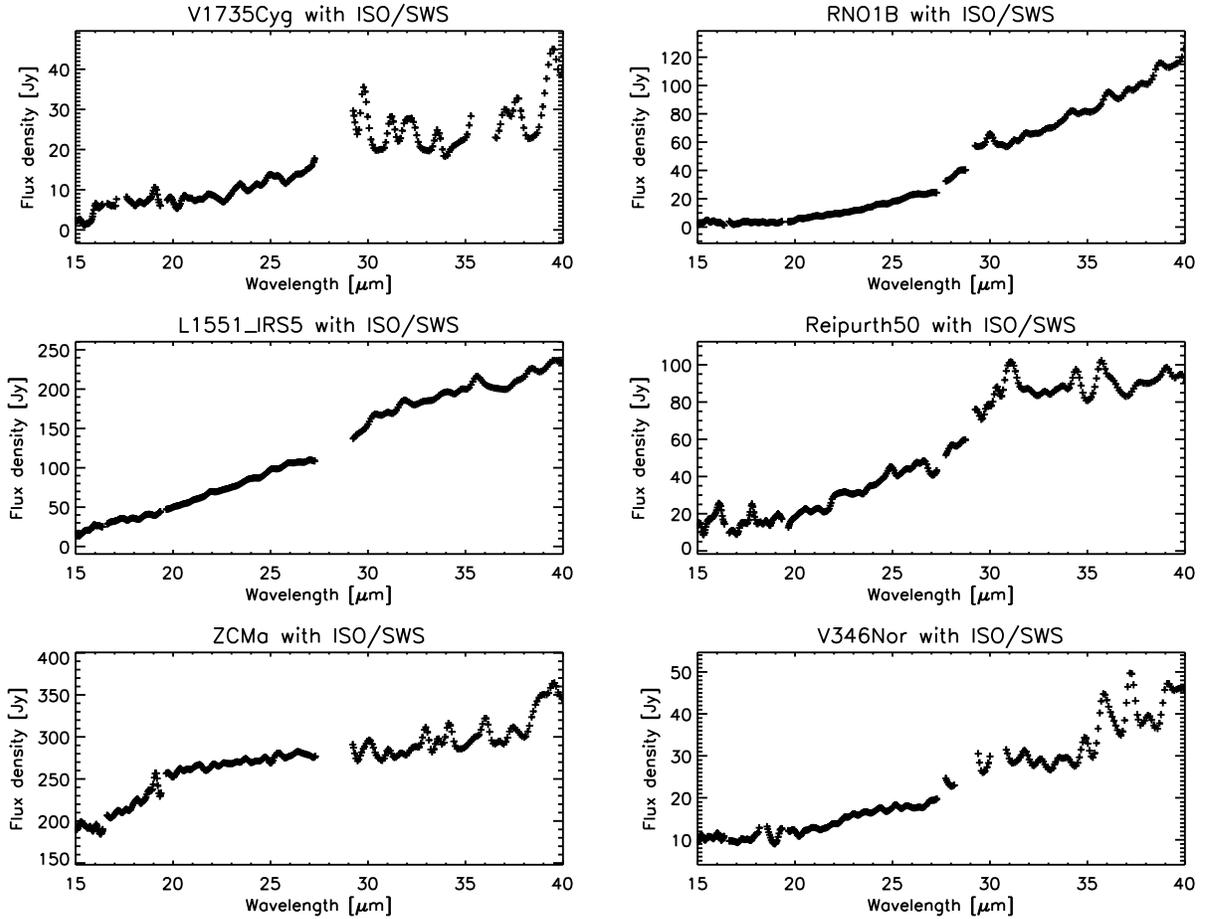}
   \caption{Same as Figure~\ref{iso_short1} but between 15 and 40\,$\mu$m. All spikes seen in the data are artefacts 
   and not believed to be real, as they showed up either in the up- or in the down-scan but not in both.}
 \label{iso_long1}
\end{figure}

\begin{figure}
\centering
\epsscale{1.}
    \plotone{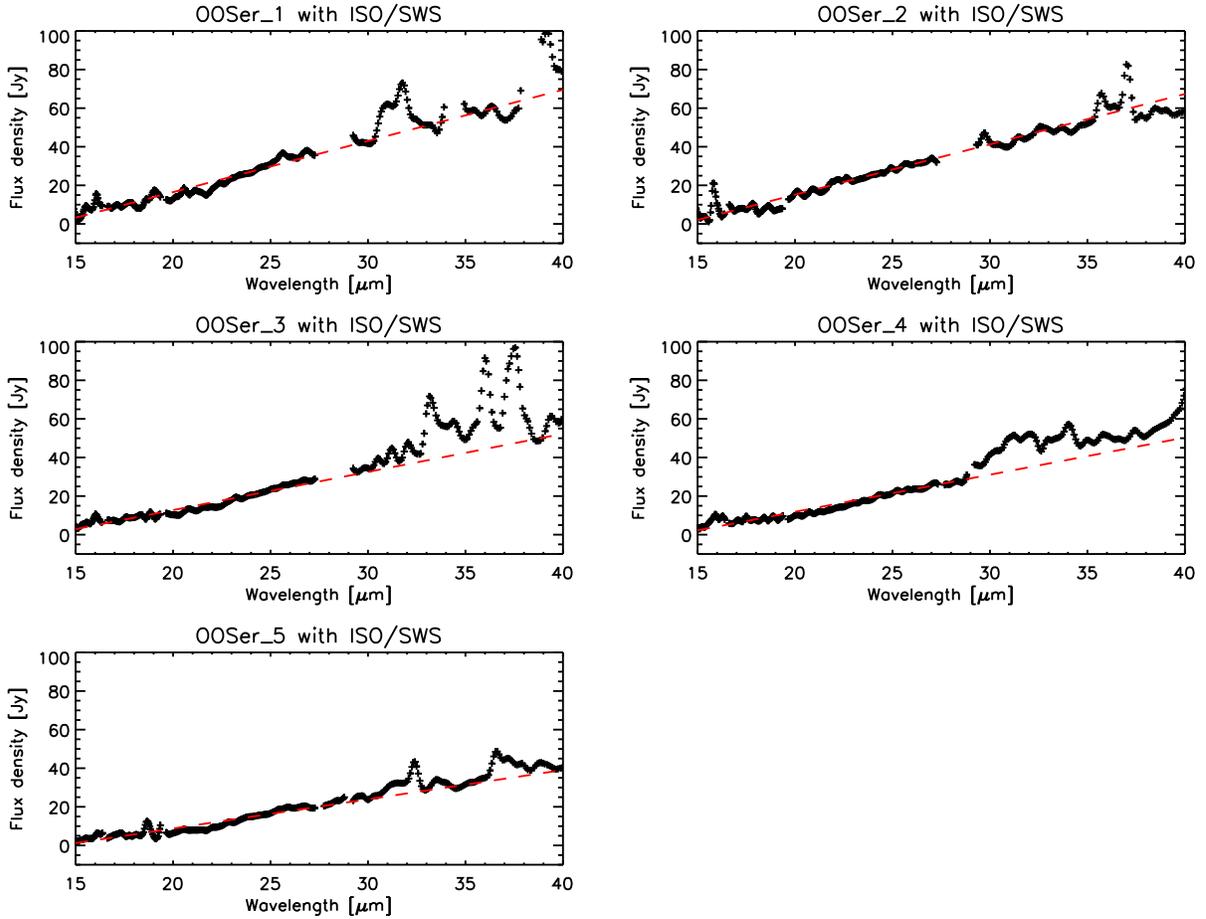}
   \caption{Same as Figure~\ref{iso_short2} but between 15 and 40\,$\mu$m. The red, dashed line illustrates 
   a fit to the data between 15 and 30\,$\mu$ to estimate the change of the slope and flux density levels during
   the decay of OO Ser (see text section 3.4.). All spikes seen in the data are artefacts 
   and not believed to be real, as they showed up either in the up- or in the down-scan but not in both. }
 \label{iso_long2}
\end{figure}

\begin{figure}
\centering
\epsscale{1.}
    \plotone{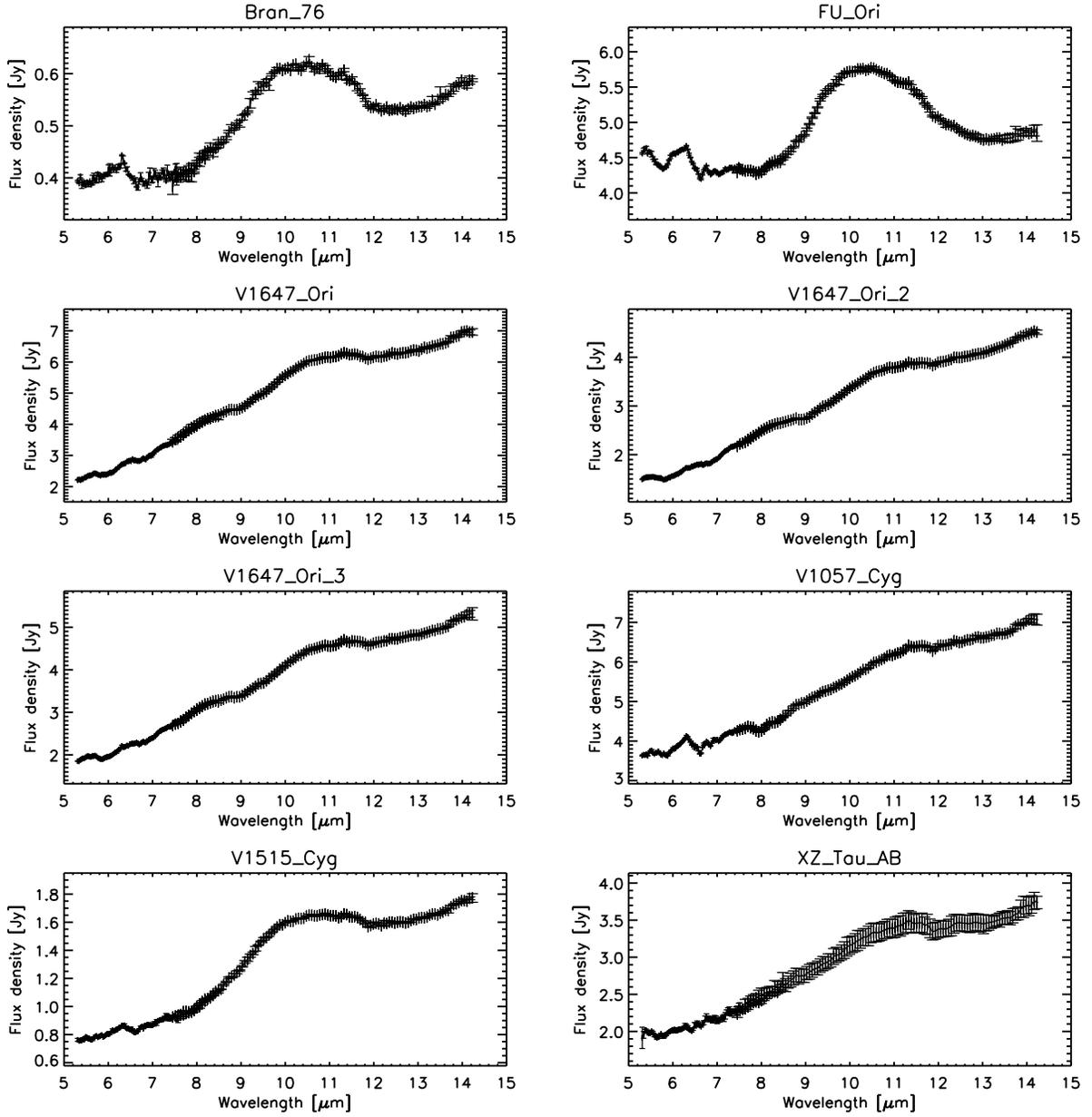}
   \caption{{\sc Spitzer} low-resolution spectra between 5 and 14\,$\mu$m.}
 \label{spitzer_short1}
\end{figure}

\begin{figure}
\centering
\epsscale{1.}
    \plotone{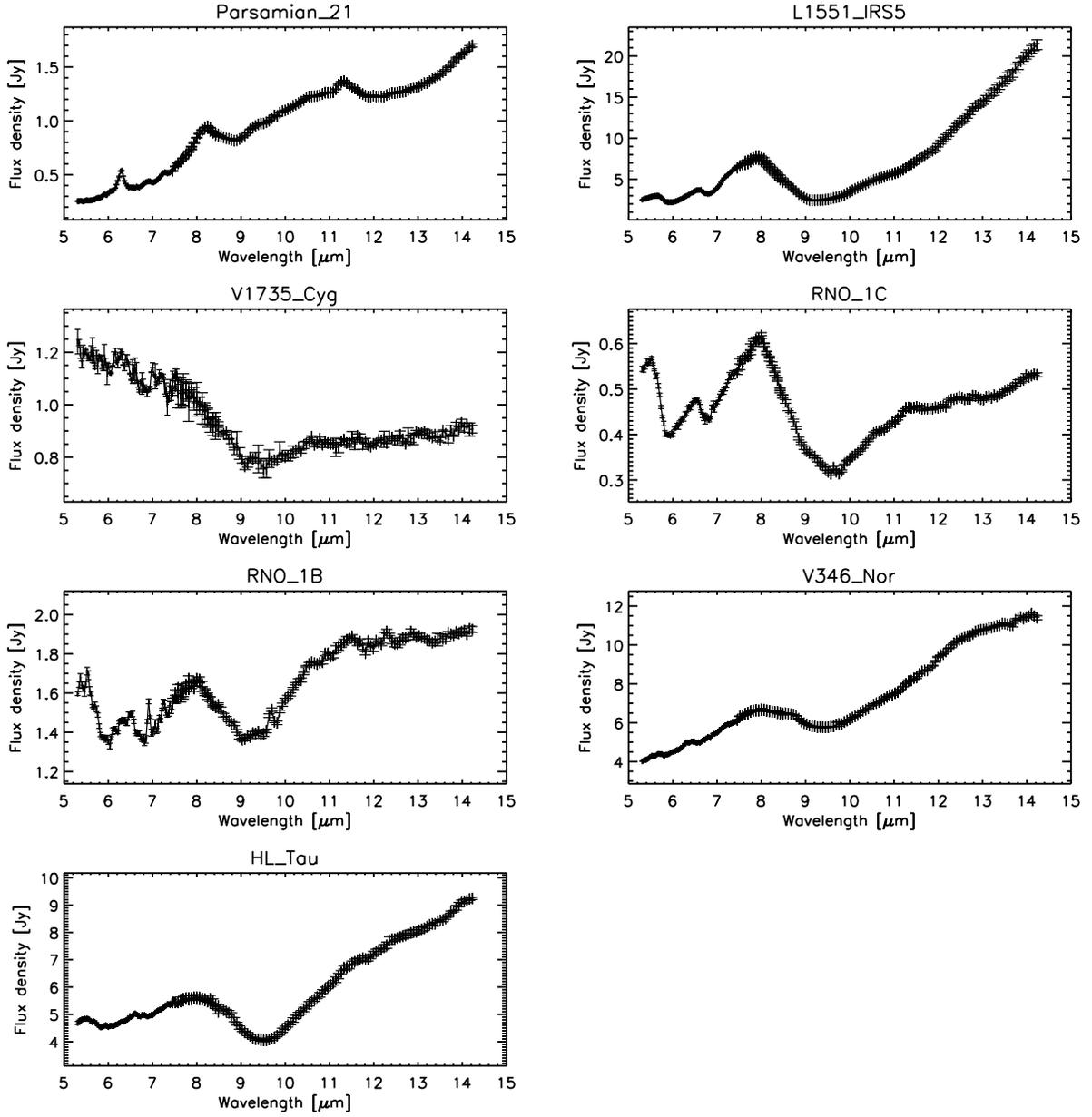}
   \caption{Same as Figure~\ref{spitzer_short1}.}
 \label{spitzer_short2}
\end{figure}

\begin{figure}
\centering
\epsscale{1.}
    \plotone{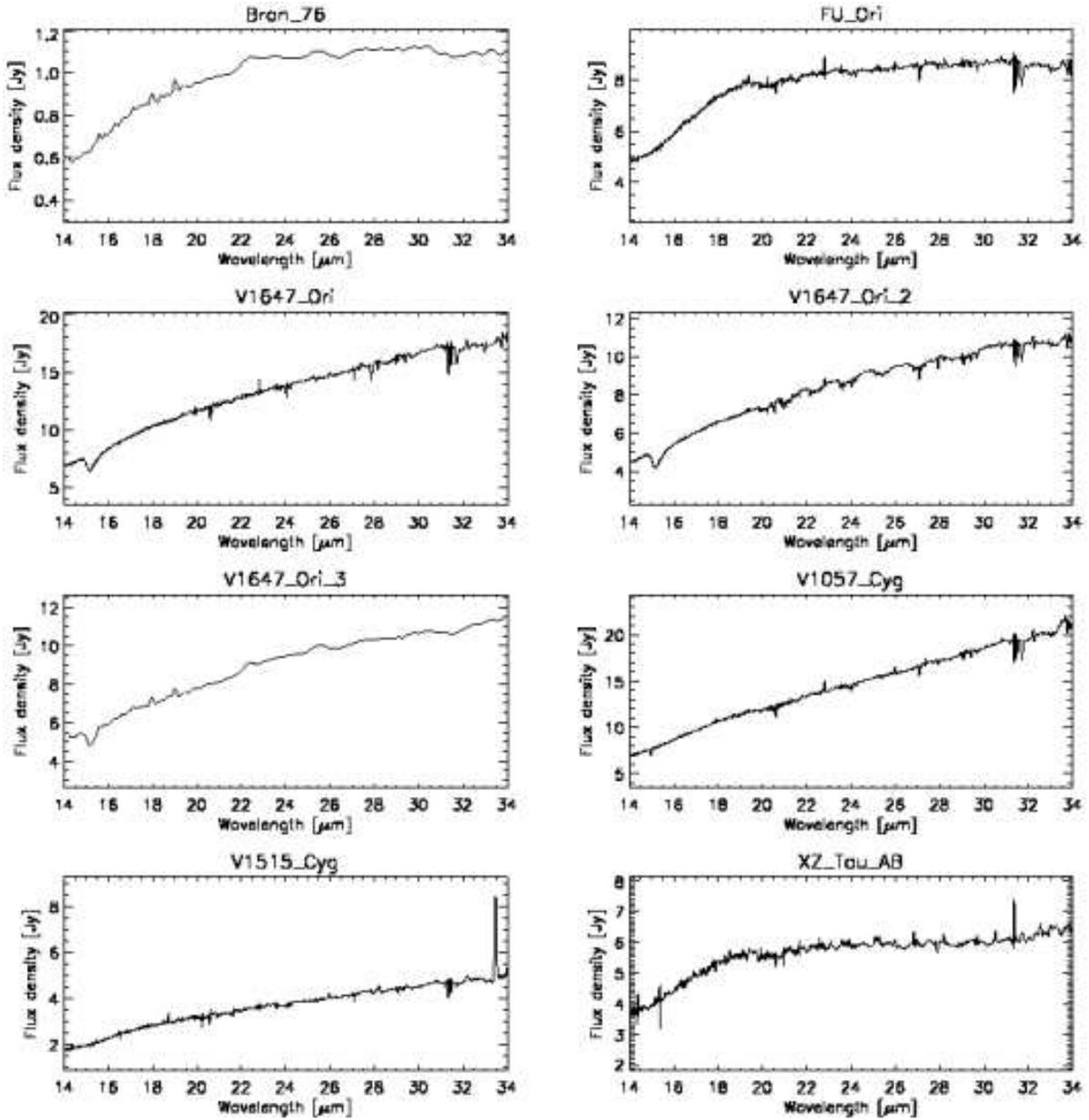}
   \caption{Same objects as in Figure~\ref{spitzer_short1}, but now showing the high-resolution 
   part of the spectrum between 14 and 34\,$\mu$m. The spectrum of V1515 Cyg shows a prominent emission line
   longward of 33\,$\mu$m which is discussed in section 3.5. Between 31 and 32\,$\mu$m all objects 
   where high-resolution data is present show artefacts from the data redcution process.}
 \label{spitzer_long1}
\end{figure}

\begin{figure}
\centering
\epsscale{1.}
    \plotone{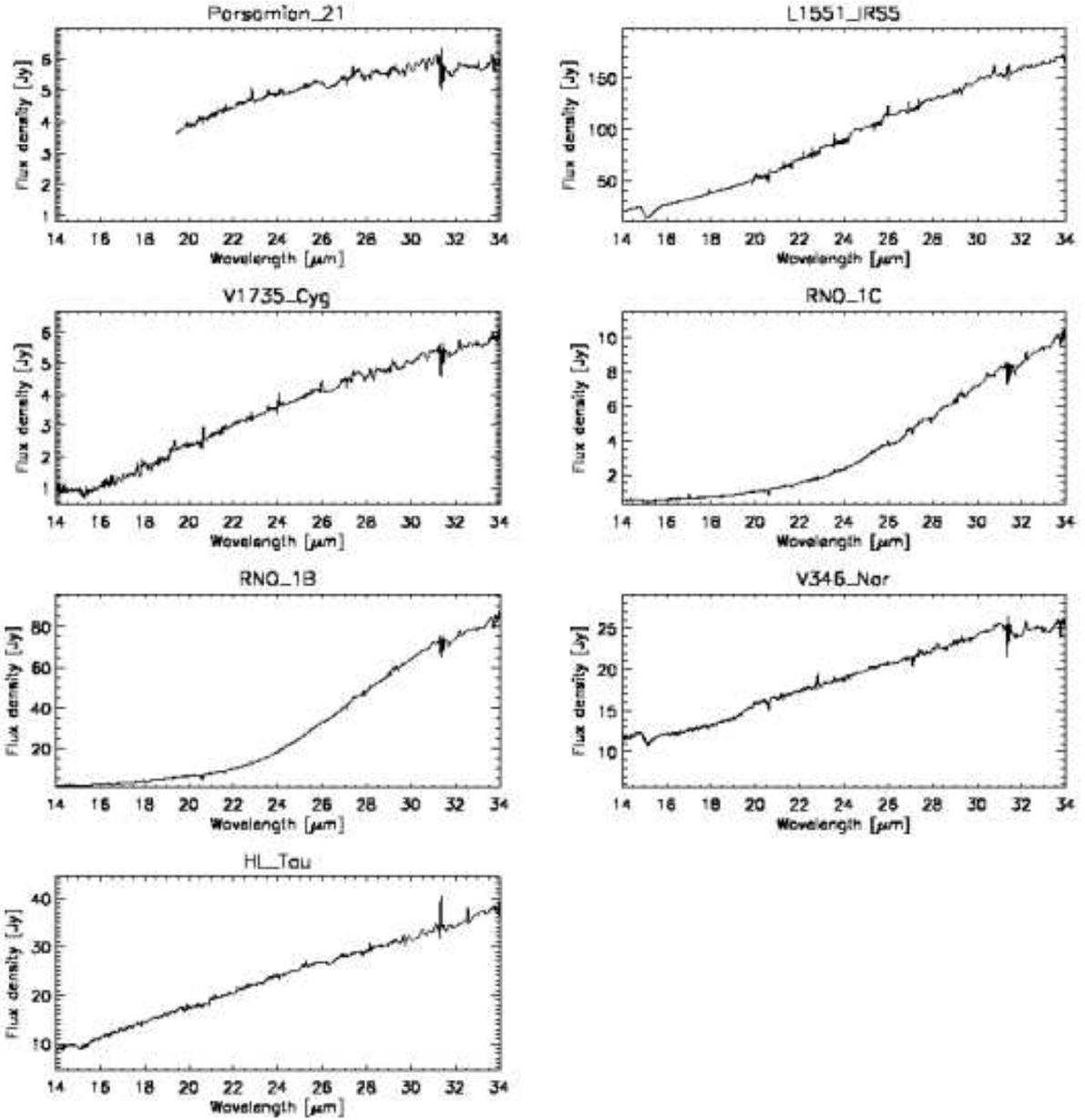}
   \caption{Same objects as in Figure~\ref{spitzer_short2}, but now showing the high-resolution 
   part of the spectrum between 14 and 34\,$\mu$m. Between 31 and 32\,$\mu$m all objects 
   where high-resolution data is present show artefacts from the data redcution process.}
 \label{spitzer_long2}
\end{figure}

Figures~\ref{iso_short1}-\ref{spitzer_long2} show the complete sample of spectra. For 
Bran 76 and the third observation of V1647 Ori only low-resolution {\sc Spitzer} data were available. 
To increase the signal to noise, all {\sc Spitzer} LH data and the SH data of V1735 Cyg were smoothed by a factor 
of three. Most of the yet remaining spikes in these spectra are not real but rather flux jumps between the 
different orders of the spectrographs.
As mentioned in the introduction of this paper, parts of the data shown here were already published: \citet{green2006} presented the {\sc Spitzer} data for 
FU Ori, V1515 Cyg, V1057 Cyg, Bran 76, and V346 Nor and used disk-envelope models to 
explain the SEDs. The spectrum of HL Tau was shown by \citet{bergin2005}. 
\citet{schuetz2005} used the ISO data for Z CMa and V346 Nor to compare with their ground-based data.
Finally, \citet{larsson2000} showed the ISO/SWS SEDs of the outbursting object OO Ser and used the ISO/SWS data
for RNO1B, Z CMa, V1057 Cyg, and V1735 Cyg for comparison. 

\begin{figure}
\centering
\epsscale{1.}
    \plotone{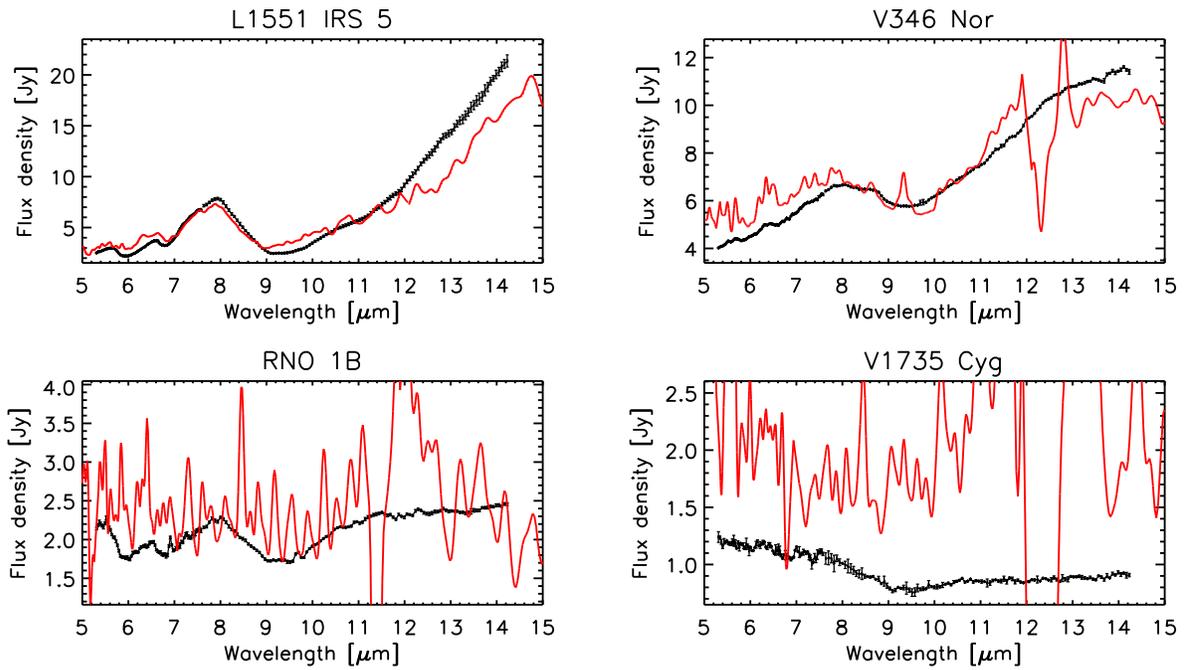}
   \caption{Comparison between ISO/SWS and {\sc Spitzer}/IRS data for objects observed with both satellites.
   The ISO data (red, solid lines) are significantly more noisy. In our sample 
   only data for objects with high flux levels or long integration times are suitable for quantitative 
   analyses.  
    }
 \label{compare_spectra}
\end{figure}

Unfortunately, for most objects the quality of the ISO/SWS data is significantly worse than that of the {\sc Spitzer} observations. Even after the data have been rebinned to a spectral resolution of 100, artefacts, i.e. 
potential emission features that were only detected either during the up- or the down-scan, remain  
in the spectra. 
In Figure~\ref{compare_spectra} we show data for four objects that were observed with both {\sc Spitzer} and ISO.  In particular, for the objects with lower flux levels the noise in the SWS data is significant. 
The reason for this is the short integration time for most objects which is reflected in the {\tt speed} parameter in Table~\ref{isojournal}. In consequence, the ISO data are mainly used for qualitative statements rather then 
for quantitative analyses throughout the rest of the paper. Only the ISO data for Z CMa (with a high flux level) and Reipurth 50 (with long integration time) will be examined in more detail in one of the consecutive sections. 
The data on Reipurth 50 is published for the first time.

From Figures~\ref{iso_short1}-\ref{spitzer_long2} it becomes clear that for all objects the flux densities
increase toward longer wavelengths indicative of warm dusty material surrounding all objects. However, Figures~\ref{iso_short1},~\ref{iso_short2},~\ref{spitzer_short1}, and~\ref{spitzer_short2} show that there are 
striking differences within the group of FUors: While some objects show a silicate emission feature in
the 10\,$\mu$m region, other objects show deep absorption profiles. Since other spectral features 
do further support such a differentiation we will in the following distinguish the objects via the behavior 
of their 10\,$\mu$m feature and discuss the two categories separately in the following subsections.
A complete overview of the most prominent spectral features between 3 and 16\,$\mu$m is given in Table~\ref{features}. If we count in V883 Ori, which was observed by \citet{schuetz2005}, and disregard 
XZ Tau and HL Tau, then 9 FUors show the silicate feature in absorption while 6 FUors show silicate emission.


\subsection{Objects with 10\,$\mu$m emission}\label{10muem}

\subsubsection{Qualitative analysis of the 10\,$\mu$m region}
\begin{figure}
\centering
\epsscale{.85}
    \plotone{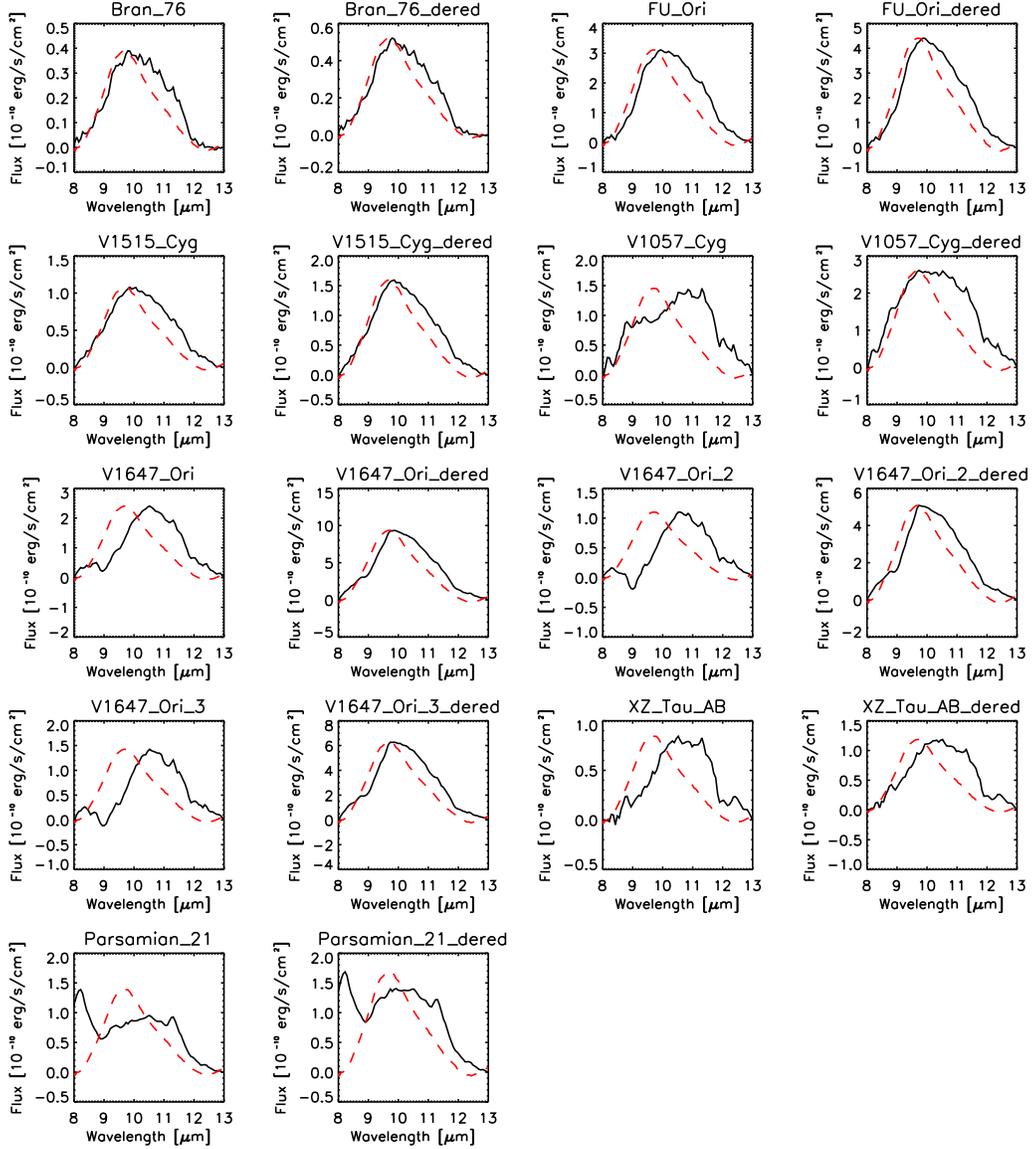}
   \caption{Continuum subtracted silicate emission profiles (black, solid lines). 
   Each spectrum is shown twice, i.e., with and without being corrected for interstellar extinction as explained 
   in the text. The assumed values for A$_{\rm V}$ are: 2.2 mag \citep[Bran 76; ][]{reipurth2002},
   2.4 mag \citep[FU Ori; ][]{skinner2006}, 3.2 mag \citep[V1515 Cyg; ][]{herbig1977}, 
   3.2 mag \citep[V1057 Cyg; ][]{herbig1977},
   11.0 mag \citep[V1647 Ori; ][]{muzerolle2005}, 1.4 mag \citep[XZ Tau; ][]{white2001}, 
   4.0 mag \citep[Parsamian 21; ][]{staudeneckel1992}. 
   We used the extinction law from \citet{mathis1990} with R$_{\rm V}$=3.1.
   For comparison the red, dashed line shows the typical ISM silicate feature observed 
   toward the galactic center from \citet{kemper2004}. 
   As mentioned in the text V1647 Ori was observed at three different epochs. 
   }
 \label{emission}
\end{figure}

In Figure~\ref{emission} we compiled all objects showing signs of silicate emission in the 10\,$\mu$m band.
To subtract the underlying continuum we fitted a polynomial of first order to the flux at 8 and 13\,$\mu$m\footnote{For Parsamian 21 we had to shift the left point of the fit to 6.6\,$\mu$m due to the special shape of the 
spectrum (see below).}. 
As we are thus far only interested in a qualitative comparison among the various objects,
the exact shape of the continuum is not important and higher oder polynomial fits did not alter the 
results relative to each other. For comparison we overplot the silicate emission feature of typical interstellar medium (ISM) dust grains, scaled to the observed spectra \citep[red, dashed lines in Figure~\ref{emission}; ][]{kemper2004}. We note that the spectrum of each object is shown twice: 
While the first plot shows the observed spectrum of the source, the second one shows
the dereddened spectrum. For the dereddening we compiled optical extinction values from the literature (see caption of Figure~\ref{emission}) and used the extinction law described in \citet{mathis1990} to derive corresponding extinction values for the MIR. It is assumed that all of the extinction is caused by material in the line
of sight toward the objects and that no self-shadowing effect (e.g., by an inclined accretion disk with a certain flaring angle) is present. The continuum was fitted for the dereddened spectra separately.
It shows that for most objects the extinction is not negligible and that
it can have substantial influence on the shape of the silicate feature (see, e.g., V1057 Cyg and V1647 Ori). For a fair comparison it thus seems reasonable to evaluate the dereddened and
not the observed spectra. 

At first glance, when comparing the spectra, it is noteworthy that all emission features differ from the typical 
shape of the typical ISM dust feature. \citet{green2006} stated that the emission peak for the 
first four objects in Figure~\ref{emission} was close to 9.5\,$\mu$m and that the dust features appeared to be 
pristine\footnote{It should be mentioned that \citet{green2006}
determined the underlying continuum from a fit to the 6\,-\,8\,$\mu$m region and not as we did from fitting a 
straight line between 8 and 13\,$\mu$m.}. We find that all features peak longward of 9.7\,$\mu$m \citep[the typical peak position for ISM type dust; ][]{kemper2004} and that they show additional flux excess compared to the ISM feature at even longer wavelengths. This indicates that dust grain processing has already set in. 
Furthermore, it should be noted that the spectrum of Parsamian 21 looks significantly different compared to the others. The most prominent characteristic
are strong emission bands around 8.2\,$\mu$m and probably also around 11.3\,$\mu$m from 
polycyclic aromatic hydrocarbons (PAHs). \citet{polomski2005} already suspected the existence of PAH emission in the spectrum of this source based on ground-based observations, but
a firm confirmation was thus far lacking. We discuss the spectrum of Parsamian 21 in more detail below. 

\begin{figure}
\centering
\epsscale{1.}
    \plotone{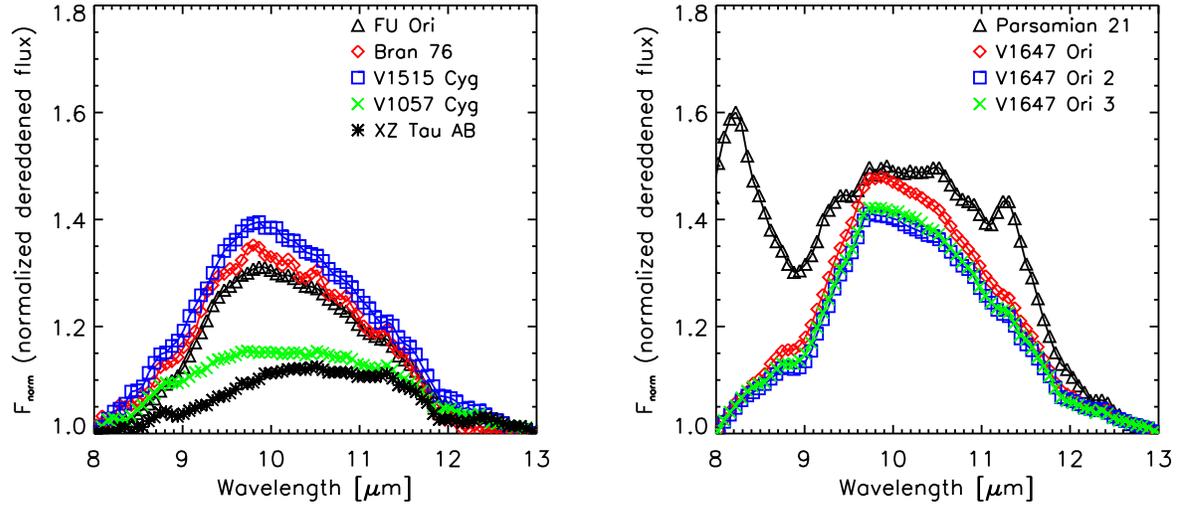}
   \caption{Dereddened and normalized 10\,$\mu$m dust emission features of the objects shown in Figure~\ref{emission}. For a better comparison the dereddened spectra are shown.}
 \label{normalizedemission}
\end{figure}

For a better comparison of the emission features, we plotted in Figure~\ref{normalizedemission} the normalized 
fluxes of the objects between 8 and 13\,$\mu$m. Following \citet{vanboekel2005}, the normalization was done via
\begin{equation}
F_{\rm norm}(\lambda)=1+\frac{F_{\rm obs}(\lambda)-F_{\rm cont}(\lambda)}{<F_{\rm cont}>}
\end{equation}
where $F_{\rm obs}(\lambda)$ is the observed flux, $F_{\rm cont}(\lambda)$ is the continuum flux and $<F_{\rm cont}>$
denotes the mean value of the underlying continuum in the considered wavelength regime. This normalization ensures 
that the shape of the emission feature is preserved. From Figure~\ref{normalizedemission} it becomes clear that the 
emission features of Bran 76, FU Ori, and V1515 Cyg are quite similar in terms of shape and strength.
The features of V1057 Cyg and XZ Tau are far less pronounced and much broader\footnote{In contrast to our analysis \citet{green2006} found the feature of V1057 Cyg to be comparable in shape (and thus in dust composition) to those of Bran 76, FU Ori, and V1515 Cyg.}. The emission profiles of V1647 Ori and Parsamian 21 are slightly stronger than those of the other objects and, as mentioned above, the latter object is the 
only one showing strong PAH emission bands. The three epochs of data for V1647 Ori allow us to study the
variability of this object in the 10\,$\mu$m region over a period of approximately five months. Figure~\ref{normalizedemission} shows that between October 2004 (epoch 1) and mid of March 2005 (epoch 2) the flux level decreased significantly. At the third epoch (end of March 2005) the flux appears to have 
slightly increased again (see also Figure~\ref{spitzer_short1}). The overall shape of the feature during the six months period did, however, not change.

\begin{figure}
\centering
\epsscale{1.}
    \plotone{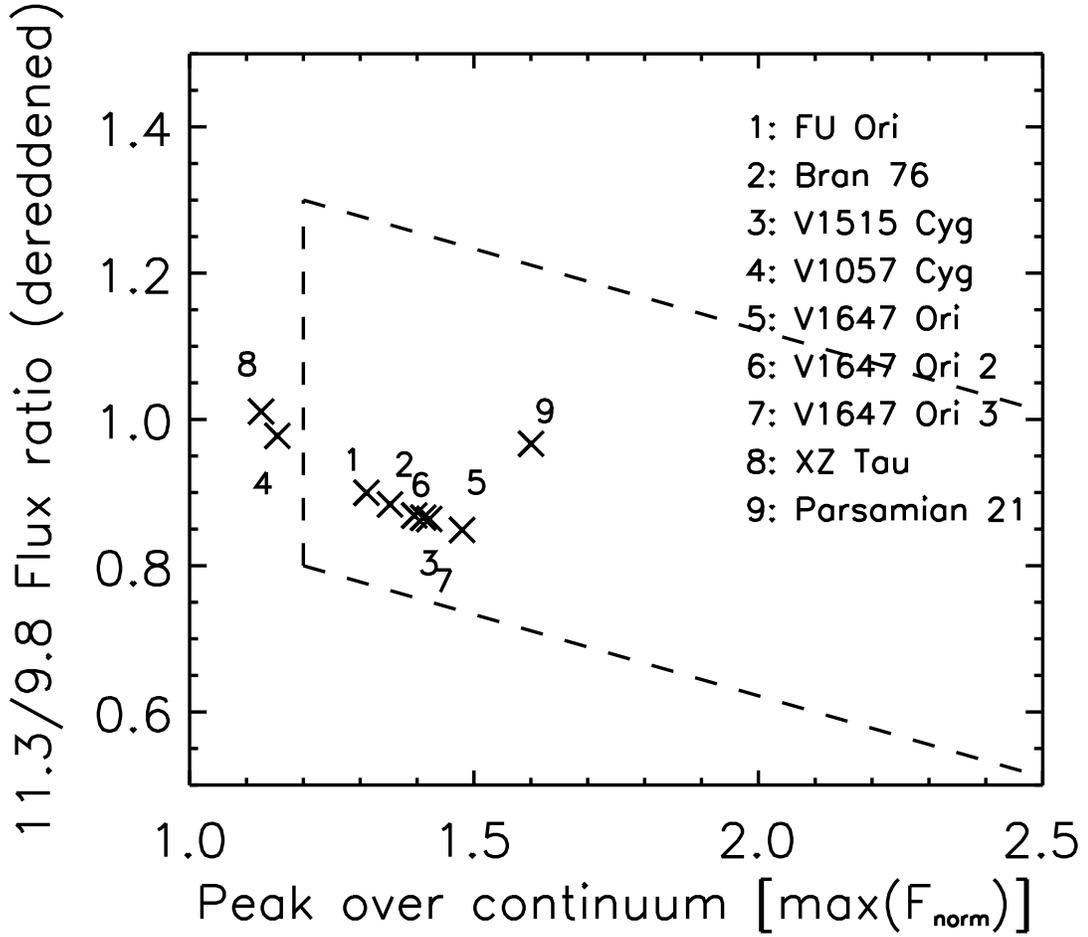}
   \caption{The flux ratio between 11.3 and 9.8\,$\mu$m plotted against 
   the normalized peak over the continuum.
   The dashed line indicates the region where typically TTauri stars \citep{przygodda2003} and Herbig Ae/Be stars 
   \citep{vanboekel2003,vanboekel2005} are located.}
 \label{fluxratio}
\end{figure}

To put these results in a broader context with other young objects and to get a first idea
on the dust grain properties, we plot in Figure~\ref{fluxratio} the flux ratio at 11.3 and 9.8\,$\mu$m against
the computed peak over the continuum in the normalized spectra, i.e., the maxima of equation (1). 
This figure also shows the region typically occupied by young TTauri stars and the slightly 
more massive Herbig Ae/Be stars 
\citep[see, e.g.,  ][]{przygodda2003,vanboekel2003,vanboekel2005,kessler2006}. We find that the objects presented here tend to have in general a relatively
weak peak over continuum emission and a flux ratio between 0.8 and 1.0. While the strength of the 
peak over the continuum is interpreted as a tracer for grain sizes (with higher peak values denoting smaller grains), the flux ratio is a more general tracer for grain processing, i.e., grain growth as well as crystallization. 
This is explained by the fact that not only the growing of grains leads to a broader and flatter silicate 
feature between 9.8 and 12.0\,$\mu$m \citep{bouwman2001}, but also the crystallization process introduces
distinct emission peaks of forsterite and enstatite longward of 10\,$\mu$m \citep[see, e.g., ][]{vanboekel2005}. 
Taking these considerations into account, we find that our sample shows clear evidence for grain growth and dust processing. For a more quantitative analysis of the dust composition we fitted a dust model to the data as 
explained in the following subsection.

\subsubsection{Dust composition}
\begin{figure}
\centering
\epsscale{1.}
    \plotone{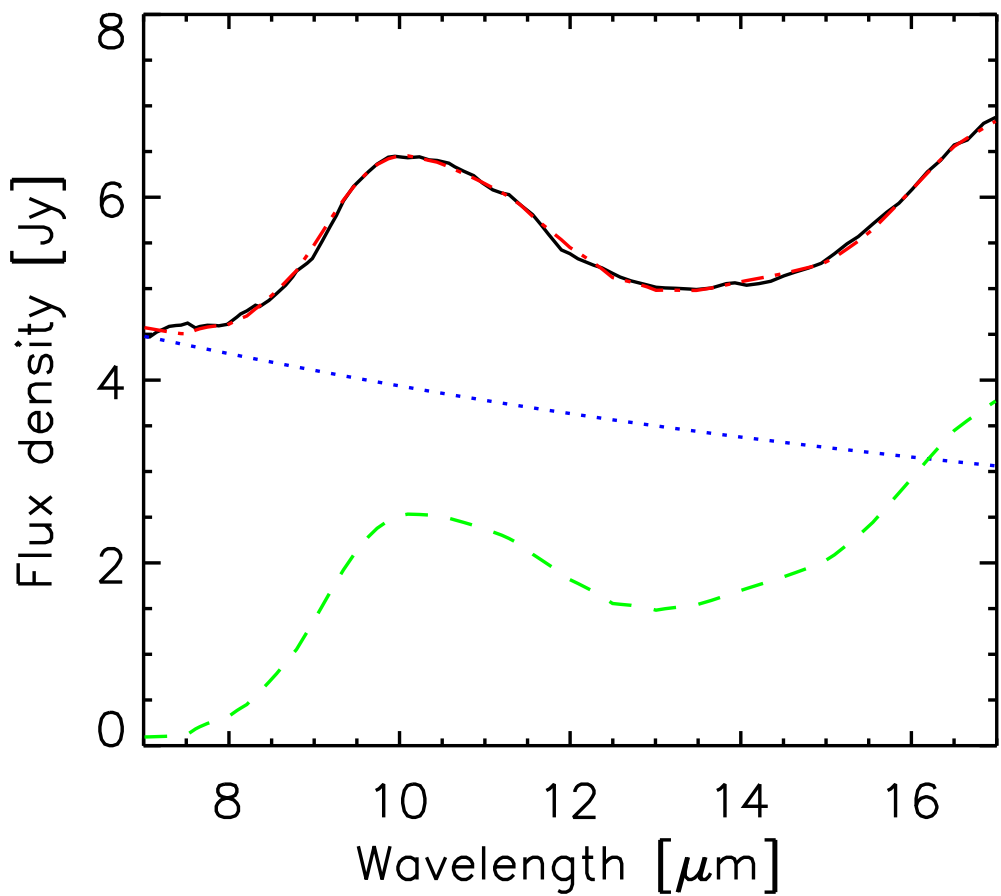}
   \caption{Results from a dust model fit to the 7-17\,$\mu$m region of the spectrum of FU Ori 
   using the model described in the text. The observed spectrum 
   is shown in the black, solid line, the 
   blue, dotted line illustrates the contribution from the fitted continuum, and the green, dashed line
   denotes the computed emission feature. The sum of the fitted 
   components is shown in the red, dash-dotted line, matching the observed spectrum.
   The mass fractions of the fitted dust species are given in Table~\ref{dust_table}. 
   The spectrum is fitted solely with amorphous grains, part of which have grown significantly.}
 \label{dust_model}
\end{figure}

In Figure~\ref{dust_model} we fitted an analytical dust model to the spectrum of FU Ori between 7 and 17\,$\mu$m.
To equally weight the low- and high-resolution part of the spectrum, we rebinned the data to a spectral resolution 
of 100, still sufficient to disentangle contributions from different dust species.
Instead of fitting a two component model to the data, with one temperature for the underlying continuum and
one temperature for the emission layer, as we did in \citet{quanz2006}, we use a new model 
described in Juh\'asz et al. 2007 (to be submitted). In this model the temperatures for 
the disk midplane, the disk surface layer and an inner disk rim are fitted by powerlaw distributions instead of  
single black body temperatures. In this respect, the model represents more realistically the temperature distribution
in a protoplanetary disk. The input parameters for the model were taken from the simple accretion 
disk model in \citet{quanz2006}. Table~\ref{temperature_table} summarizes the fitted parameters of the individual
temperature components.

The dust model was based on opacities for six dust species (amorphous olivine and pyroxene, crystalline forsterite 
and enstatite, silica, and amorphous carbon) in three grain sizes (0.1, 1.5, and 6.0\,$\mu$m). 
References for the optical properties of the grains are given in Juh\'asz et al. 2007 (to be submitted) and
\citet{quanz2006}. In contrast the analysis presented in \citet{quanz2006}, 
this time we included also carbon grains in the fit. However, as shown in Figure~\ref{dust_model} and summarized in Table~\ref{dust_table}, the best fit ($\chi^2$\,=\,45.3) does not require any carbon grains. The derived mass fractions agree well with the results based on the more simple two component dust model used in \citet{quanz2006}. 
While a large fraction of big, amorphous grains contribute to a high-temperature continuum, the smaller
grains are responsible for the observed emission feature. Still, we emphasize that, according to our analysis, 
also in the disk surface layer large grains are present: Comparing the derived composition to that of the interstellar medium \citep{kemper2004,vanboekel2005}, 
we find a significantly higher fraction of 1.5\,$\mu$m grains in the 
dust of FU Ori. Furthermore, the model does predict a negligible amount of crystalline grains ($<$\,0.01), no silica dust and, as mentioned above, no carbon grains. Certainly, the values for the derived mass fractions should not be taken literally, as, for instance, the temperature structure in the disk and the dust 
opacities are not perfectly known. However, the models presented here and in \citet{quanz2006} 
agree on two very important aspects: (1) there is 
clear evidence for dust coagulation in the spectra of FUors, and (2) there is hardly any
contribution from crystalline grains detectable in the data.
This lack of crystalline grains was already noticed and discussed in \citet{quanz2006}, but we will
reinvestigate this issue below.

\subsubsection{The PAH features of Parsamian 21}\label{par21}
\begin{figure}
\centering
\epsscale{1.}
    \plotone{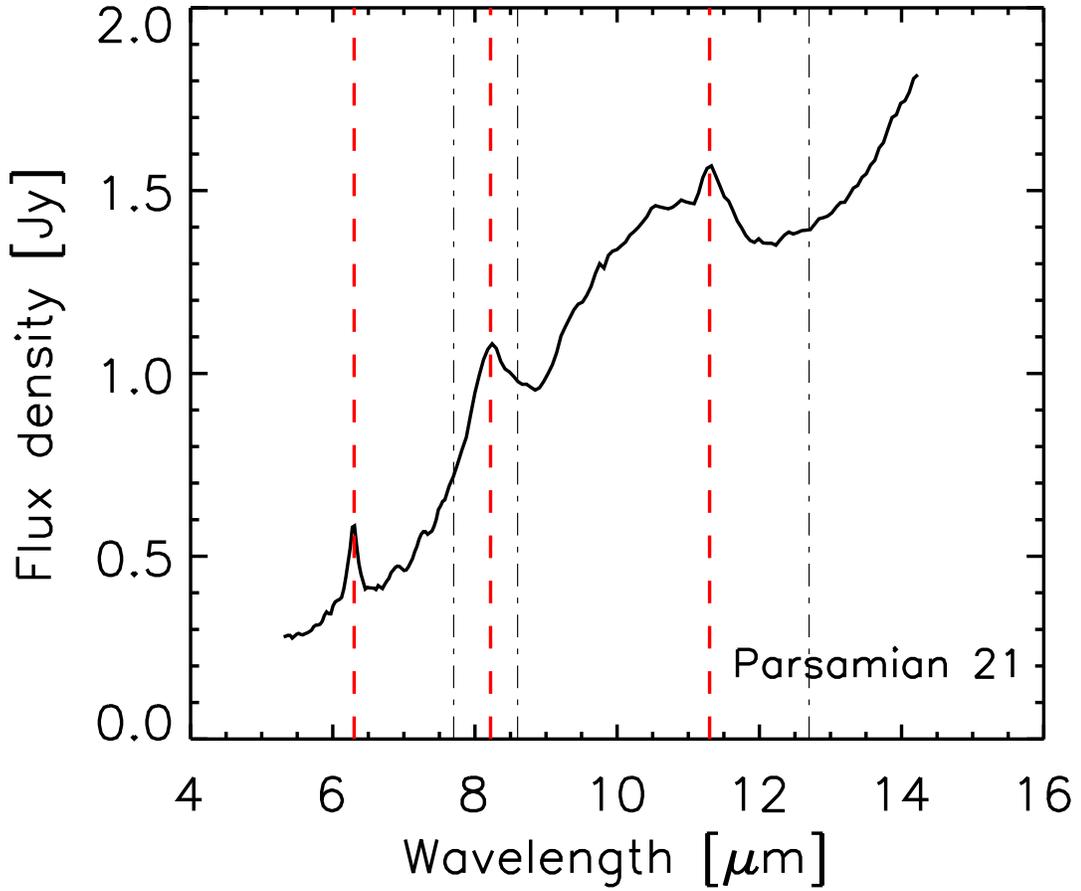}
   \caption{Dereddened {\sc Spitzer} MIR spectrum of Parsamian 21 (assuming A$_{\rm V}$\,=\,4.0 mag)
   illustrating the strong PAH emission features not seen
   in any other FUor. Note also the steeply rising continuum and the underlying silicate dust emission.
   The vertical lines indicate the positions of typical PAH emission bands: strong emission is detected
   at 6.3, 8.2 and 11.3\,$\mu$m (red, dashed lines), no emission is seen at 7.7, 8.6 and 12.7\,$\mu$m.
   }
 \label{parsamian_pah}
\end{figure}

As its features are significantly different than those of the other FUors, we plot in Figure~\ref{parsamian_pah}
the dereddened spectrum of Parsamian 21 and analyze it in more detail. 
The vertical lines illustrate, where typically PAH emission bands are located. While prominent PAH emission is present at 6.3, 8.2 and (possibly) 11.3\,$\mu$m, no bands are seen at 7.7, 8.6 and 12.7\,$\mu$m. In addition, the spectrum bears signs of underlying silicate emission and it is difficult to 
disentangle in the 11.3\,$\mu$m emission band the possible contributions from crystalline forsterite and 
PAHs. 

PAHs are commonly seen in the spectra of HAeBe stars \citep[e.g.,][]{meeus2001,acke2004,sloan2005} and, more
recently, PAH emission was also detected in TTauri spectra \citep{geers2006}. However, the 
spectrum of Parsamian 21 differs significantly from the PAH spectra of other young objects: it shows 
a broad and strong feature at 8.2\,$\mu$m which is barely detected and the main characteristic of 
group C objects in \citet{peeters2002}. These objects show no PAH complex 
at 7.7\,$\mu$m and 8.6\,$\mu$m but the peculiar broad band at 8.2\,$\mu$m whose origin yet remains unknown \citep{peeters2002}. In section~\ref{par21_discussion} we
will critically review the findings presented here and discuss their implications with respect to the 
FUor status of Parsamian 21.

\subsubsection{Qualitative analyses of the 5\,-\,8\,$\mu$m region}
\begin{figure}
\centering
\epsscale{1.}
    \plotone{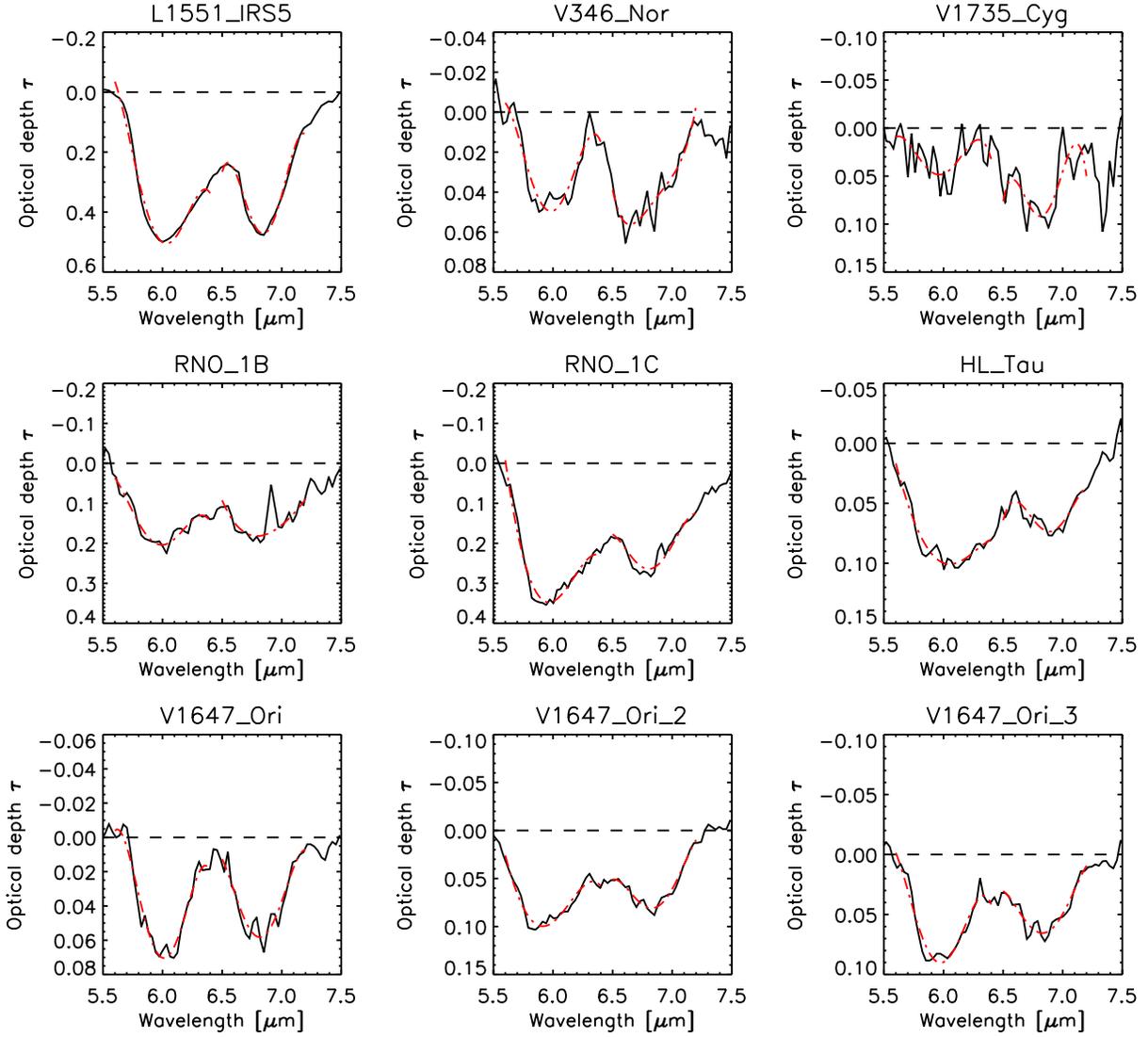}
   \caption{The 6.0 and 6.85\,$\mu$m ice bands observed toward some of our objects (black lines). Overplotted
   are the polynomial fits of fourth order (red, dashed-dotted line) to determine the optical depths of the
   absorption bands. }
 \label{absorption6}
\end{figure}

As already noted and explained 
by \citet{green2006} FU Ori, Bran 76, V1515 Cyg, and also V1057 Cyg show absorption bands of gaseous
H$_2$O at 5.8 and 6.8\,$\mu$m (Figure~\ref{spitzer_short1}) 
coming from a collection of rotation-vibration bands.\footnote{Interestingly, already \citet{sato1992} found evidence for water vapor absorption between 1 and 3\,$\mu$m in the spectra of FU Ori, V1515 Cyg, and V1057 Cyg, 
but also L1551 IRS 5, V1057 Cyg, and Z CMa.}
V1647 Ori, however, shows 
evidence for ice absorption bands at 6.0 and 6.85\,$\mu$m (see also Figure~\ref{absorption6}). 
We mention already here that this objects shows an additional ice feature at 15.2\,$\mu$m 
due to CO$_2$ (Figure~\ref{absorption15}).
V1647 Ori is the only object in the sample showing 
the 10\,$\mu$m silicate feature in emission accompanied by significant ice absorption bands. A more detailed 
analysis of these ice features is given in section~\ref{10muabs}. Finally, as we will discuss in the following section, Parsamian 21 shows clear evidence for a PAH emission feature at 6.2\,$\mu$m.


\subsection{Objects with 10\,$\mu$m absorption}\label{10muabs}
\subsubsection{Analyses of the 10\,$\mu$m region}
\begin{figure}
\centering
\epsscale{1.}
    \plotone{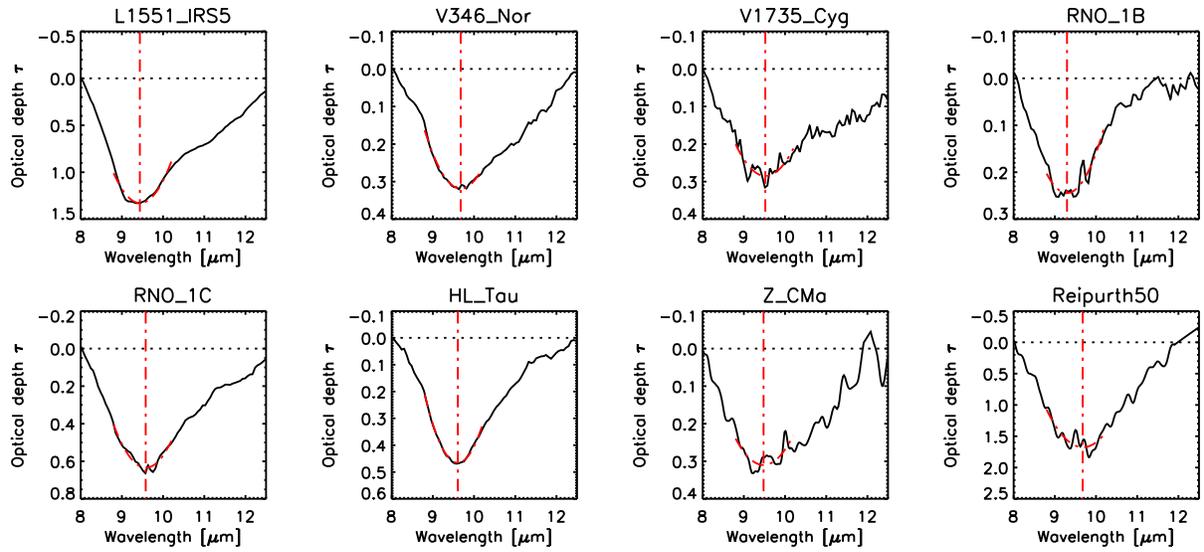}
   \caption{Observed optical depths of the 10\,$\mu$m silicate absorption feature (black, solid line). 
   The dotted line shows the
   assumed continuum. The vertical 
   red, dash-dotted lines indicate the position of the maximum optical depth of the second order polynomial fitted 
   to the data between 8.8 and 10.2 (also plotted in red, dash-dotted lines).}
 \label{absorption}
\end{figure}

In Figure~\ref{absorption} we plot the optical depth in the 10\,$\mu$m region for the objects showing the silicate  band in absorption. To derive the optical depth, we fitted the continuum with a polynomial of first order to the observed spectra between 8 and 13\,$\mu$m, assuming that no absorption due to silicates is present at these wavelengths. Also here the exact shape of the continuum is difficult to estimate and any fit always carries
uncertainties. However, even fits with higher order polynomials did not alter the results significantly and the 
main conclusions remained valid. Based on the assumed continuum we then computed the optical depth $\tau$ using  
\begin{equation}
F_{\rm obs}(\lambda)=F_{\rm cont}(\lambda)\cdot e^{-\tau}
\end{equation}
where $F_{\rm obs}(\lambda)$ is the observed flux and $F_{\rm cont}(\lambda)$ the assumed continuum. 
To derive the wavelength position and value of the maximum optical depth, we did the following:
As some spectra are quite noisy (e.g., V1735 Cyg, RNO 1B, Z CMa, and Reipurth 50) we fitted the data 
between 8.8 and 10.2\,$\mu$m, i.e., the region with the greatest optical depth, with a polynomial of second order
(shown as red, dash-dotted lines in Figure~\ref{absorption}).
The wavelength position of the maximum optical depth (vertical, red, dash-dotted lines in Figure~\ref{absorption}) 
and the corresponding value are then derived from the fit. 
From the spectra with a high signal to noise ratio and strong absorption profiles (e.g., the {\sc Spitzer} spectra of L1551 IRS 5 and HL Tau) it becomes clear that the fit matches the shape of the absorption feature quite well and that 
this approach seems reasonable. The black, dotted horizontal line in Figure~\ref{absorption} shows the 
assumed continuum level. In Table~\ref{extinctiontable} we summarize the derived maximum optical depth for each object $\tau_{\rm silicate}$ and the corresponding
wavelength. We give also values for the extinction in the silicate band A$_{\rm silicate}$ and 
estimates for the corresponding extinction in the optical A$_{\rm V}$. A$_{\rm silicate}$ was derived via
\begin{equation}
A_{\rm silicate}=-2.5\,{\rm log_{10}}\,({\rm e}^{-\tau{_{\rm silicate}}})\quad .
\end{equation}
The errors for
$\tau_{\rm silicate}$ and A$_{\rm silicate}$ are based on the 1-$\sigma$ uncertainties in the spectra
and the resulting fluctuations of the polynomial fit. 
A$_{\rm V}$ and its error are derived from averaging over the extinction models of \citet{mathis1990} and \citet{weingartnerdraine2001} and assuming two different values of R$_{\rm V}$ for each model (R$_{\rm V}$\,=\,3.1 and 5.0 for \citet{mathis1990} and R$_{\rm V}$\,=\,3.0 and 5.5 for \citet{weingartnerdraine2001}). 
As already noted by \citet{cohenkuhi1979} one has to be careful when deriving a value for the optical extinction based on the silicate feature, as uncertainties in the underlying continuum and unknown dust compositions 
influence the results. Hence, for comparison, Table~\ref{extinctiontable} gives also values for A$_{\rm V}$ found in the literature. One has to keep in mind, though, that also here certain assumptions have been made and 
partly different observing techniques were applied potentially leading to discrepancies in the derived
values for A$_{\rm V}$.

For V346 Nor and RNO 1C the values for A$_{\rm V}$ agree quite well. Also for Z CMa, Reipurth 50 and L1551 IRS 5
the values are in general agreement. Due to the high extinction toward the latter two objects,
a good estimate of A$_{\rm V}$ is difficult. We would like to point out, however, that a value of
A$_{\rm V}$ $\gtrsim$\,150 mag for L1551 IRS 5 as found by \citet{campbell1988} and mentioned, for instance, by \citet{stocke1988}, \citet{white2000} and \citet{fridlund2002} 
appears to be an overestimate. In yet unpublished NIR data taken
with the {\sc Omega2000} camera at the Calar Alto Observatory (Spain) we do see a near-infrared (NIR)
counterpart to the L1551 IRS 5 system with K$_{\rm S}$\,$\approx$\,9.7 mag. Even if this was purely scattered
light from a disk the intrinsic K magnitude of the central object(s) (it might be a multiple system) can not
be lower. Thus, the observed K-magnitude is a lower limit for the intrinsic K-luminosity of the source.
The above mentioned value for A$_{\rm V}$ would translate into A$_{\rm K}\approx$\,15 mag, and thus 
the absolute K-band magnitude of L1551 IRS 5 would be $\approx -11$ mag for an assumed distance of 140 pc. 
For young low-mass objects this appears to be definitely too luminous even if accretion luminosities 
are taken into account as well. For V1735 Cyg and
HL Tau our derived values for A$_{\rm V}$ are lower than those found in the literature. However, for 
V1735 Cyg no errors are given for the value found in the literature. 
For HL Tau the optical extinction derived from the radiative transfer 
model by \citet{menshchikov1999} may suffer from insufficient information about the source geometry or dust opacities which can easily lead to large uncertainties in the computed figures. 
The comparison for these objects clearly shows that, in particular for embedded objects, it is difficult to derive 
consistent values for A$_{\rm V}$ if different techniques are applied.

Finally, RNO 1B also shows a discrepancy between our value
for A$_{\rm V}$ and that given by \citet{staudeneckel1991}. In this case, however, the 
silicate absorption feature bears evidence for additional superposed emission as will be described in the 
next paragraph. Thus, we attribute at least part of the missing optical depth to an underlying
silicate emission feature. 

\begin{figure}
\centering
\epsscale{1.}
    \plotone{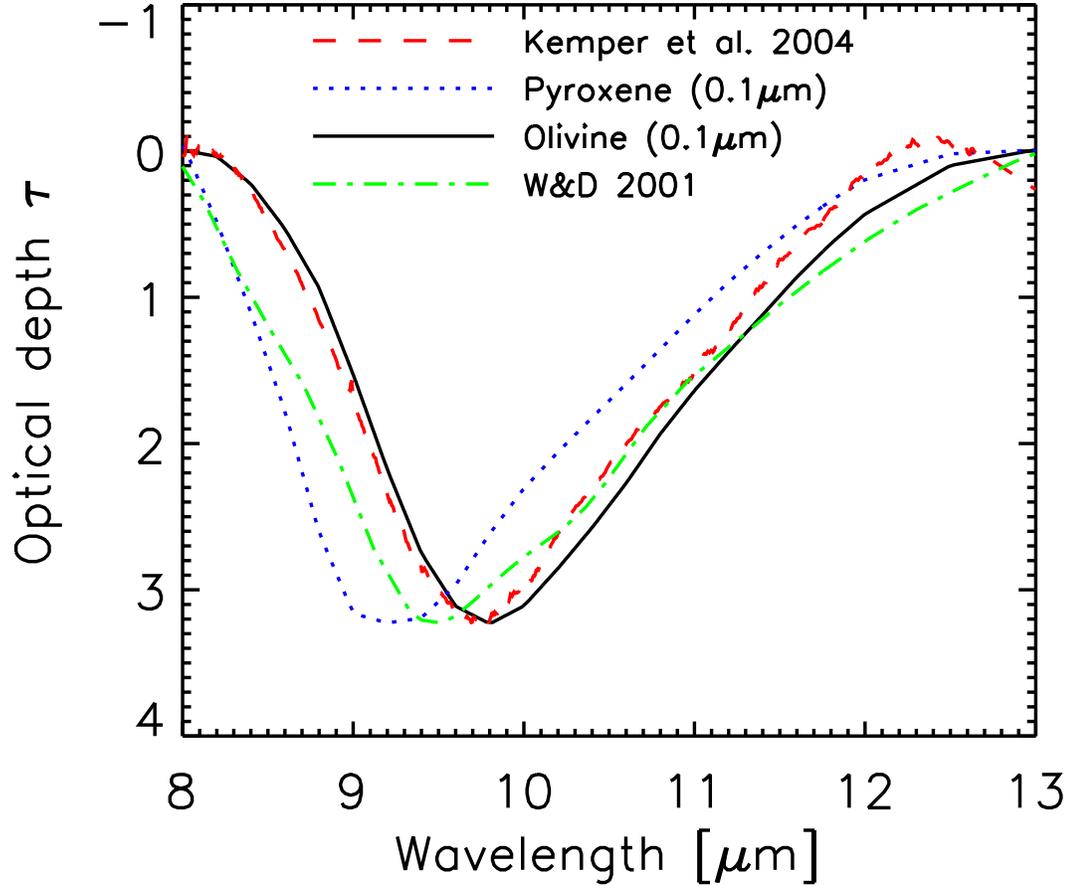}
   \caption{Optical depths for different dust grain populations on an arbitrary scale. 
   The black, solid line shows the profile for 0.1\,$\mu$m sized amorphous 
   olivine grains \citep{dorschner1995}, the red, dashed 
   line shows the ISM feature toward the galactic plane from \citet{kemper2004}, the green, dash-dotted line is the 
   profile based on the astronomical silicates from \citet{weingartnerdraine2001} and \citet{draine2003}, 
   and the blue, dotted line is for 0.1\,$\mu$m sized amorphous pyroxene
   grains \citep{dorschner1995}.}
 \label{abs_profiles}
\end{figure}

To measure not only the depth of the absorption profile, but to get also an idea of the
dust composition responsible for the observed absorption, we analyzed the shape of the 
silicate feature. In Figure~\ref{abs_profiles} we show the optical depths computed or measured for different 
dust grain compositions. Depending on the composition the wavelengths of the maximum optical depth
changes. The most extreme cases are amorphous olivine grains with a size of 0.1\,$\mu$m peaking 
at a wavelength slightly longward of 9.7\,$\mu$m, and amorphous pyroxene grains with the same size 
peaking around 9.2\,$\mu$m wavelength. As mentioned above, in Table~\ref{extinctiontable} 
we already give the wavelengths
of the maximum optical depths observed toward our sources and it shows that there is indeed a broad 
range of values, indicating different dust compositions. 

\begin{figure}
\centering
\epsscale{1.}
    \plotone{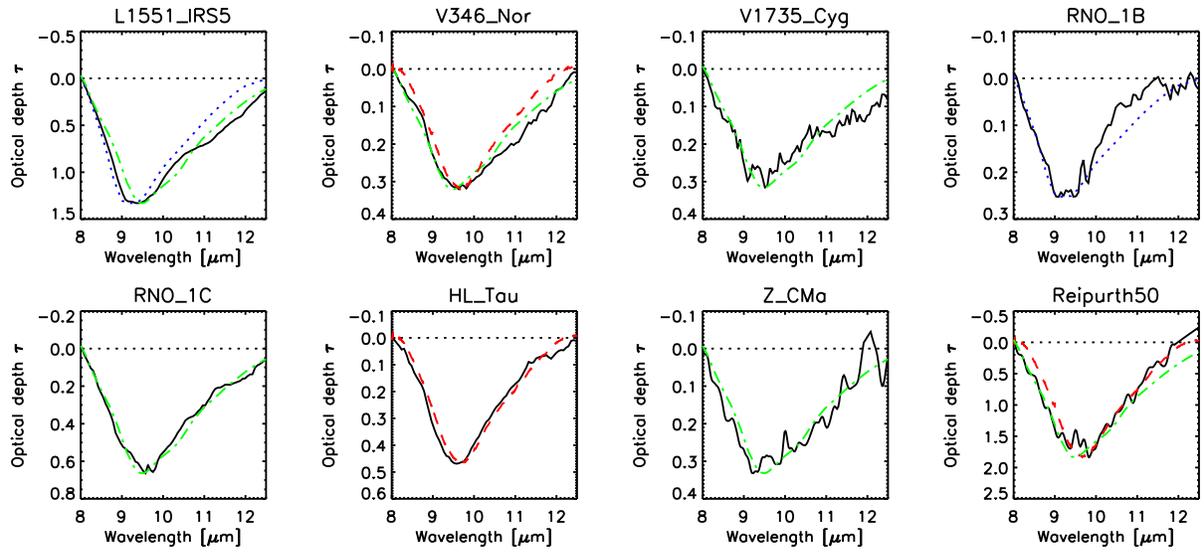}
   \caption{Same observed spectra as in Figure~\ref{absorption} (black, solid lines),  
   but now overplotted with reference spectra with different dust compositions (same color and
   line style code as in Figure~\ref{abs_profiles}). }
 \label{absorption2}
\end{figure}

In Figure~\ref{absorption2} we plot again the observed optical depth of our objects 
as shown in Figure~\ref{absorption}, but now we overplot one or two of the absorption profiles presented
in Figure~\ref{abs_profiles}, scaled to the maximum observed optical depth. Whether this scaling
is appropriate is difficult to determine but it shows that most observed features agree
reasonably well with one or two of the reference profiles. Three objects (L1551 IRS 5, V346 Nor and V1735 Cyg) 
show additional absorption longward of 11\,$\mu$m when compared to the dust features, but at least part
of this absorption can be attributed to H$_2$O libration bands \citep[e.g., ][]{pontoppidan2005}.
V1735 Cyg, RNO 1C and Z CMa seem to agree best with 
the dust model computed by \citet{draine2003} based on the particle size distribution from \citet{weingartnerdraine2001}. 
V346 Nor requires shortward of 10\,$\mu$m a mixture of 
the same model and the dust composition observed toward the galactic center \citep{kemper2004}.
Longward of 10\,$\mu$m the former model alone fits better. For Reipurth 50 a combination 
of both models explains the data well over the whole wavelength regime with the \citet{draine2003} model
fitting the short and the \citet{kemper2004} model fitting the long wavelength part.
L1551 IRS 5 is best fitted with a mixture of small, amorphous pyroxene grains and the model from \citet{draine2003}.
The profile of HL Tau agrees extremely well with the galactic center dust profile from \citet{kemper2004}.
To account for a small shift toward shorter wavelengths a little bit more amorphous pyroxene seems to be required 
or small uncertainties in the position of the peak absorption from the continuum fit can explain this shift. 
Finally, the absorption profile of RNO 1B is special in two ways: First, the 
short wavelength range is best fitted solely with small, amorphous 
pyroxene grains. And secondly, between 10 and 11.5\,$\mu$m
the profile shows a significantly lower optical depth than any of the reference profiles. In combination with 
our findings for the visual extinction (see above) this suggests that RNO 1B shows silicate emission 
superposed on the absorption feature. Interestingly, the apparent decrease in optical depth longward of 10\,$\mu$m 
indicates that at these wavelengths the contribution of the emission is relatively larger compared to the 
other wavelengths in the silicate band. This in turn means that the emission profile does not
have the typical shape of the ISM dust feature but a broader and 
more evolved profile like the other FUors 
emission profiles shown in Figure~\ref{emission}.

\subsubsection{The 15.2\,$\mu$m CO$_2$ ice feature}
\begin{figure}
\centering
\epsscale{1.}
    \plotone{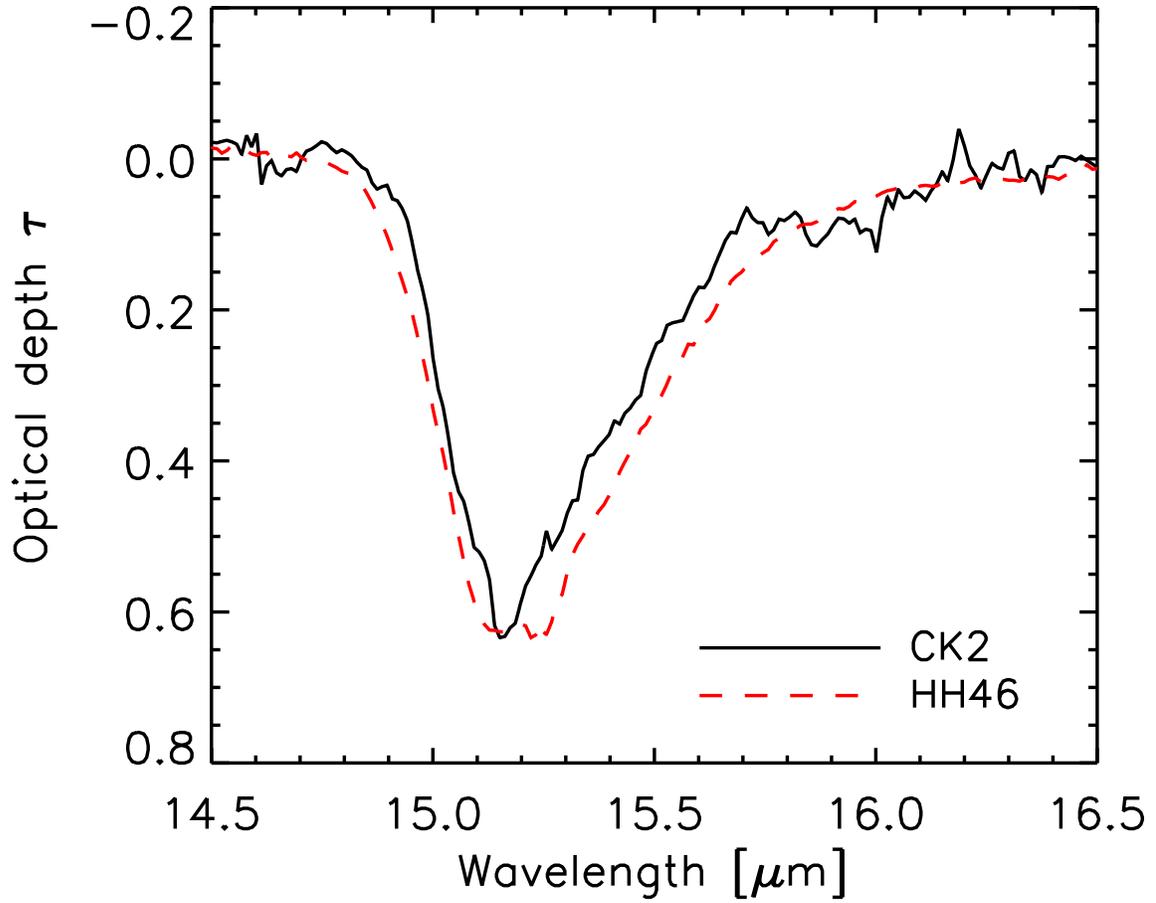}
   \caption{Comparison of the 15.2\,$\mu$m CO$_2$-ice feature between the heavily extincted background 
   object CK2 \citep{knez2005} and the embedded young source HH46 \citep{boogert2004}. Due to
   higher temperatures in the vicinity of the embedded protostar crystallization occurs and 
   and a double-peaked sub-structure appears characteristic of the pure CO$_2$ matrix \citep{ehrenfreund1998}.}
 \label{absorption15_comparison}
\end{figure}

\begin{figure}
\centering
\epsscale{1.}
    \plotone{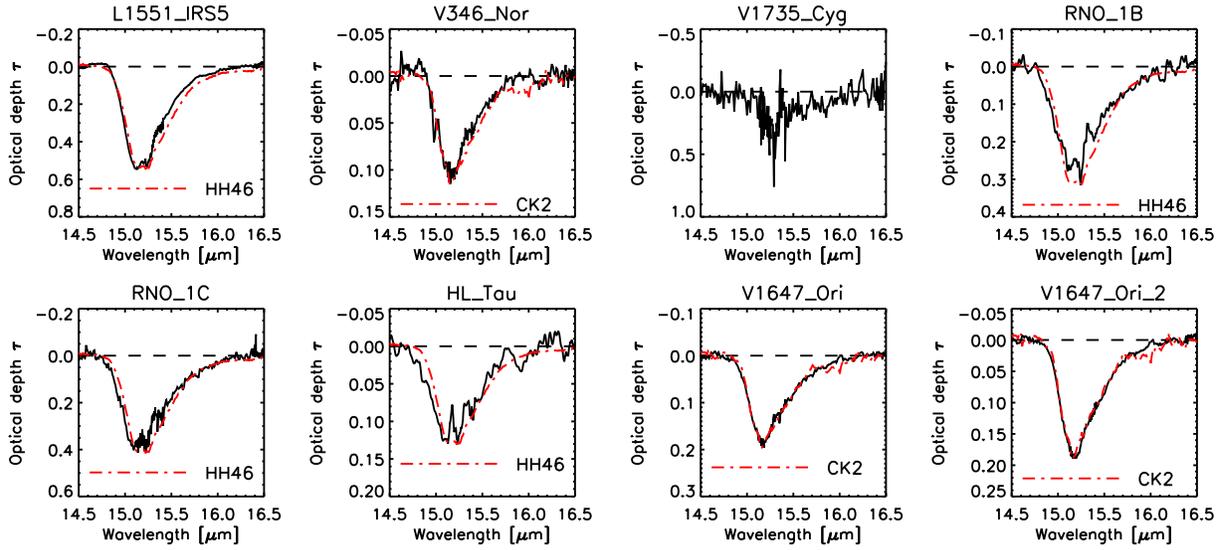}
   \caption{The 15.2\,$\mu$m CO$_2$-ice feature observed toward some of our objects (black, solid lines).  
    The spectra of HL Tau, RNO 1B and V1735 Cyg were smoothed by a factor of 20 to increase the signal to noise. 
    Overplotted is either the scaled spectrum of HH46 or CK2 (red, dashed-dotted line; see
    Figure~\ref{absorption15_comparison}) depending on which profile fits better to the FUor data.
    Due to the remaining high noise level in the spectrum of V1735 Cyg no clear comparison to 
    either reference object was possible. The absorption feature of V1647 Ori observed at the third 
    epoch is not shown as these are only low-resolution data, while the data presented here (first and second epoch)
    were taken with the high-resolution spectrograph.}
 \label{absorption15}
\end{figure}

The shape of the bending mode profile of CO$_2$ ice around 15.2\,$\mu$m does not only allow 
a detailed analysis of the involved ice inventory, but it enables us to derive 
information about potential ice processing due to heating effects \citep[e.g.,][]{pontoppidan2005}. 
In Figure~\ref{absorption15_comparison} we present two reference cases which we will use in the following
as a benchmark for comparison with the FUor spectra. The first object in Figure~\ref{absorption15_comparison}
(CK2) is a highly extincted background star behind the Serpens dark cloud \citep{knez2005}, and the second object (HH46) is an embedded low-mass protostar \citep{boogert2004}. 

While the spectrum of CK2 probes mainly the ice composition of the 
intervening dark cloud, the spectrum of HH46 bears information about the immediate environment of the young 
star. Apart from being slightly broader, the spectrum of HH46 shows a double-peaked sub-structure which 
is caused by crystallization and effective segregation of the  CO$_2$ and H$_2$O ice involved in the absorption.
These processes take place when ice mixtures with concentrations of  CO$_2$/H$_2$O$\ge$1 are heated \citep{boogert2004}. In space, this phase transition from amorphous to crystalline ice is expected to occur between 50-90 K and thus at higher temperatures than typically found in cold, dark molecular clouds. Thus, the 
15.2\,$\mu$m ice feature of HH46 shows that (at least) part of the ice must already have been heated by
the embedded protostar. In fact, \citet{boogert2004} fit the feature with a two-component ice model
based on laboratory spectra with one component being highly processed polar ice with a laboratory 
temperature of T$_{\rm lab}$\,=\,125 K, and the other component being an H$_2$O-rich, CH$_3$OH-deficient cold ice
with T$_{\rm lab}$\,=\,10 K.\footnote{The presence of CH$_3$OH in the ice feature can potentially be traced by 
the shape of the long-wavelength wing of the CO$_2$ profile showing additional absorption if CH$_3$OH 
is present in higher abundances.}
The spectrum of CK2 was fitted by \citet{knez2005} solely with cold ice
components. They used a polar mixture of H$_2$O:CO$_2$=1:1 and H$_2$O:CO$_2$=10:1 at 10 K with a ratio of 2:1 and 
an additional apolar component of CO:N$_2$:CO$_2$=100:50:20 at 30 K. The overall polar fraction was
assumed to be 78\,\%.

In Figure~\ref{absorption15} we present the 15.2\,$\mu$m features observed toward our FUor sample. 
To derive the optical depths we fitted the continuum with a straight line fixed around 14.65 and 16.3\,$\mu$m.  
For each object we overplot either the (scaled) spectrum of HH46 or CK2 depending on the shape of the profile. While
the profiles of V1647 Ori and V346 Nor agree better with the profile of CK2 (the spectrum representing
unprocessed ice), L1551 IRS 5, RNO 1B and 1C, and HL Tau show evidence for a double-peaked sub-structure 
and thus heating effects and processed ice. The comparatively bad quality of the V1735 Cyg spectrum does not 
allow a solid comparison to either reference spectrum. It is noteworthy that the spectrum of V1647 Ori 
is almost an exact copy of CK2 indicating that the ice composition is mostly identical. Based on the 
fitted reference spectrum we computed also the optical depth $\tau_{15.2\,\mu{\rm m}}$ 
for each object and summarized the results in 
Table~\ref{iceextinctiontable}. The errors are based on the 1-$\sigma$ uncertainties in the observed spectra.

\subsubsection{The 6.0 and 6.8\,$\mu$m ice features}
Although frequently observed toward high- and low-mass sources,
the two well-known ice features at 6.0 and 6.8\,$\mu$m are 
quite complex and difficult to interpret. Certainly, a large fraction of the 
optical depth of the 6.0\,$\mu$m band can be attributed to H$_2$O ice, but also other species might contribute 
to this absorption feature \citep{keane2001}. For instance, slightly shortward, at roughly 5.85\,$\mu$m, an additional absorption shoulder is sometimes superposed \citep[e.g.,][]{pontoppidan2005,keane2001} for which formaldehyde (H$_2$CO) and formic acid (HCOOH) are theoretical candidates. Without any additional information
(e.g., the 3.08\,$\mu$m band of H$_2$O or the 3.47\,$\mu$m band of H$_2$CO) it is thus difficult to 
determine the true water ice content in the 6.0\,$\mu$m band.

An additional absorption feature at 6.85\,$\mu$m is often observed toward protostars \citep[e.g.,][]{keane2001}, 
but also toward the extincted background star CK2 \citep{knez2005} and the edge-on disk CRBR 2422.8-3423 \citep{pontoppidan2005}. Although a final identification of this band has yet to be provided,
NH$_4^+$ seems to be one of the most promising candidates \citep{schutte2003,pontoppidan2005}. 
However, \cite{dishoeck2004} for example, mention also methanol (CH$_3$OH) as a potential carrier of this
absorption band. 

Given all these uncertainties we restrict ourselves in this section to the computation
of the optical depth of both of the above mentioned absorption bands. Like in the previous section
we assumed a straight line for the continuum anchored at 5.4 and 7.6\,$\mu$m.
To derive the optical depths we then fitted a polynomial
of fourth order to both absorption dips to eliminate the noise in the spectra. 
Figure~\ref{absorption6}
shows the observed spectra between 5.5 and 7.5\,$\mu$m on an optical depth scale (black lines) with the
resulting fits overplotted (red, dashed-dotted lines). The computed optical depth for each band and each 
object is given in Table~\ref{iceextinctiontable}. Like in the previous section the errors are based on the 1-$\sigma$ uncertainties in the observed spectra.

\subsection{The fading of OO Ser and V1647 Ori}
For two objects (OO Ser and V1647 Ori) we have multi-epoch data and can derive some conclusions
on the variability of these objects.
As illustrated in Figure~\ref{iso_long2}, we fitted a straight line to the data of OO Ser between 15 and 30\,$\mu$m 
to estimate the decay in flux density observed over the 5 epochs (see, Table\ref{isojournal}).
The wavelength range was chosen as in this regime all spectra are still relatively clear of artefacts and spikes partly seen at longer wavelengths. In Table~\ref{ooserflux} we summarize the flux density level at 
20 and 30\,$\mu$m at each epoch and give also the slope measured between 15 and 30\,$\mu$m. The 
errors are derived from the goodness of the fit to the data. Between the 
first observations on April 14, 1996, and the last observations, carried out September 22, 1997, the flux densities 
decreased to roughly 50\% of the initial values. These data demonstrate that OO Ser faded rapidly over relatively short timescales and that it might be an intermediate object between a typical FUor and an EXor having
fading timescales of several decades and months, respectively. Based on photometric 
monitoring at infrared wavelengths, \citet{kospal2007} came to a similar conclusion and predicted that 
OO Ser should return to its pre-outburst luminosity not before 2011. 

In Table~\ref{ooserflux} we also summarize the flux densities of V1647 Ori observed at 8, 20 and 30\,$\mu$m at three different epochs. The errors are taken directly 
 from the {\sc Spitzer} spectra. 
It shows that also this object faded significantly over  a period of less than 5 months between October 2004 and March 2005. In addition, short time variations 
in the flux levels seem to be present, as between the two epochs in March 2005 the object became slightly brighter again. In consequence, these data 
support the assumption that the outburst of V1647 Ori may also be intermediate between FUor- and EXor-type events  similar to OO Ser \citep{muzerolle2005,acosta-pulido2007}. 

\subsection{Additional emission lines}
\begin{figure}
\centering
\epsscale{1.}
    \plotone{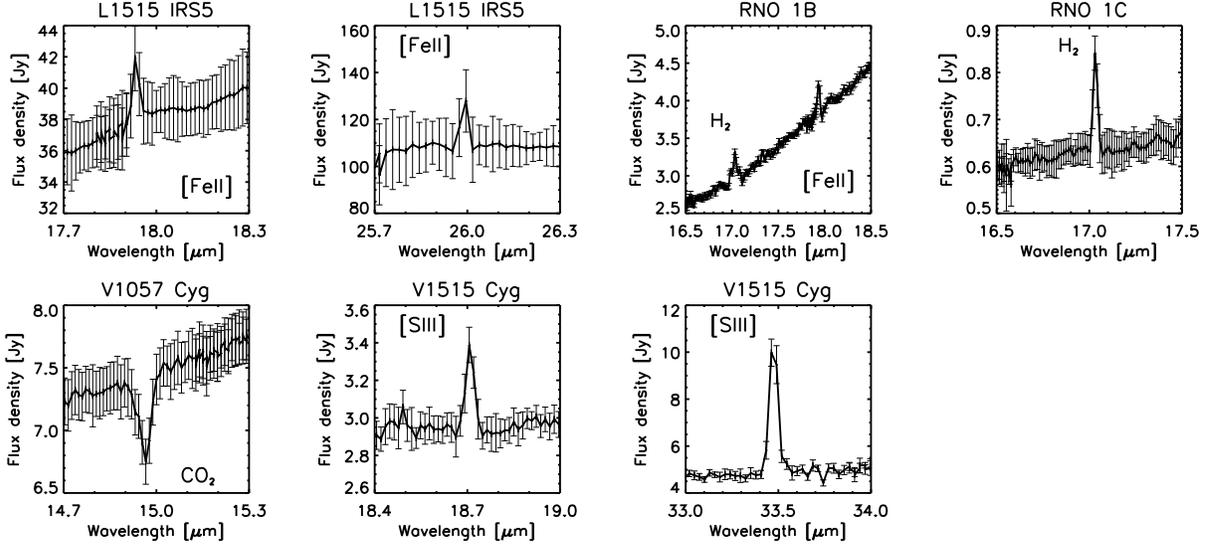}
   \caption{Zoom into emission lines and absorption lines detected in the high-resolution part 
   of the spectra of some objects (see text).}
 \label{lines}
\end{figure}

For completeness we show in Figure~\ref{lines} absorption and emission lines identified in the high-resolution regime
of the {\sc Spitzer} spectra, part of which are difficult to identify in Figures~\ref{spitzer_long1} and~\ref{spitzer_long2}. Already \citet{green2006} noted the [S III] emission lines at 18.7 and 33.4\,$\mu$m 
in the spectrum of V1515 Cyg and argued that they originate from extended emission in the region and not
from the object itself. The spectra of RNO 1B and RNO 1C show evidence for H$_2$ quadrupole emission 
around 17\,$\mu$m and additional H$_2$ lines in the low-resolution part of the spectrum at shorter wavelengths. In \citet{quanz2007rno} we analyzed these emission lines in detail and concluded that they are related to shocks within a molecular outflow powered by the nearby embedded object IRAS 00338+6312. [Fe II] lines around 17.9\,$\mu$m are
present in the spectra of L1551 IRS 5 and RNO 1B, and L1551 IRS 5 shows also 
the [Fe II] line near 26.0\,$\mu$m. The [Fe II] lines of L1551 IRS 5 were already detected in the ISO/SWS 
spectrum of this source and attributed to hot and dense material located close to the root of the outflow \citep{white2000}. The line intensities did not fit the predictions from shock models. Concerning the [Fe II]
line in the spectrum of RNO 1B, it seems likely that it arises in the outflow shocks that also excite the 
H$_2$ emission lines discussed in \citet{quanz2007rno}. 
Finally, V1057 Cyg shows gaseous CO$_2$ absorption slightly shortward of 15.0\,$\mu$m.


\section{Discussion}
\subsection{Two categories of FUors}
The results presented here suggest that the sample of FUors can be  
divided into two categories based on the observational appearance of the 10\,$\mu$m silicate feature.
We decided to call the absorption 
feature objects "Category 1"-FUors (9 objects in Table~\ref{features}) 
and the emission feature objects "Category 2"-FUors (6 objects in Table~\ref{features}). 
In the following we discuss the properties of the two 
categories in more detail.


\subsubsection{Category 1 FUors: silicate and ice absorption features}
In general it is possible to observe the silicate feature in absorption if the 
circumstellar disk surrounding a young object 
is seen close to edge-on. \citet{menshchikovhenning1997} showed that typically for a disk with an opening angle of 20$^{\circ}$ between the upper and lower disk surface the silicate feature appears 
in absorption only if the disk inclination is $\lesssim 10^{\circ}$ from edge-on.
Given the amount of objects showing silicate absorption compared to 
the number of objects with silicate emission and assuming a random distribution of the orientation
of the accretion disks, it is unlikely that all Category 1 FUors are seen edge-on. 
Rather, 
these objects are still more deeply embedded in their molecular envelopes covering a larger 
solid angle than the edge-on disk alone. These envelopes cause the ice and dust absorption features. 

Figure~\ref{absorption2} illustrates that all but one 
silicate absorption bands are fitted best either with 
the silicate composition observed toward the Galactic center by \citet{kemper2004} 
or with the astronomical silicates from \citet{weingartnerdraine2001} and \citet{draine2003}\footnote{L1551 IRS 5 requires a fraction of additional amorphous pyroxene grains.}. This finding shows that all absorption features can be fitted with small, amorphous silicates and hence the extinction is caused by pristine and not processed dust. 
Only the spectrum of RNO 1B is difficult to fit with any ISM dust compostition, but, as already mentioned, there are strong hints that the absorption feature is altered by a superposed emission feature. 

\begin{figure}
\centering
\epsscale{1.}
    \plotone{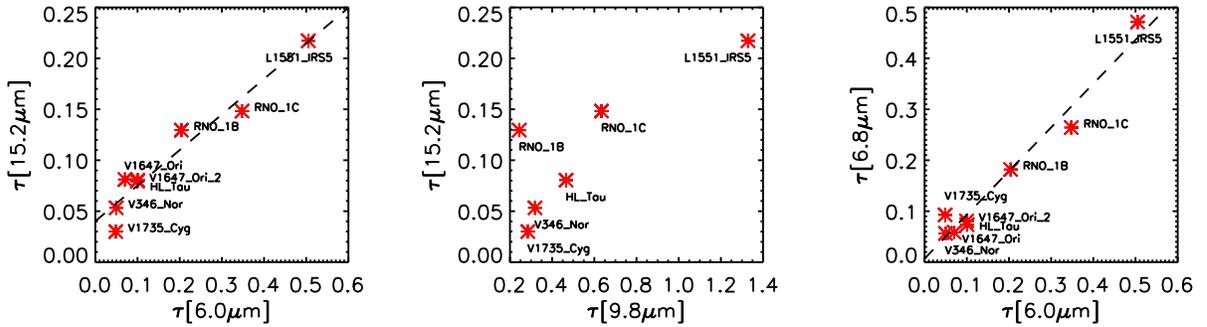}
   \caption{Correlation between the optical depths of ice features at 15.2 and 6.0\,$\mu$m (left), 
   ices and silicates at 15.2 and 9.8\,$\mu$m (middle), and ice features at 6.8 and 6.0\,$\mu$m (right).
   V1647 Ori is not shown in the middle plot as its silicate feature is seen in emission and not 
   in absorption (see text). }
 \label{absorption_correlation}
\end{figure}

In Figure~\ref{absorption_correlation} we show the observed optical depths of the ices and the silicate
feature in a scatter plot to search for any correlation. Although the range of optical depths we probe here
is limited, it seems that at least for the ices there seem to be correlations (left and right plot in 
Figure~\ref{absorption_correlation}). Fitting a straight line to the data we find:
\begin{equation}
\tau_{15.2\,\mu{\rm m}}=(0.349\pm0.037)\cdot\tau_{6.0\,\mu{\rm m}}+(0.040\pm0.009)
\end{equation}
\begin{equation}
\tau_{6.8\,\mu{\rm m}}=(0.858\pm0.067)\cdot\tau_{6.0\,\mu{\rm m}}+(0.007\pm0.016)
\end{equation}
This implies that the physical and chemical conditions within the envelopes and clouds 
causing the absorption are similar. A more detailed analysis, e.g., the determination of different 
ice abundances relative to water ice, is beyond the scope of this work. For such an analysis the
water ice feature around 3\,$\mu$m is required as it suffers less from additional contributions of other 
ice species compare to the 6\,$\mu$m feature. 

For a correlation between the optical depths of the silicates and ices the situation is slightly different because, as we have already pointed out, 
the silicate feature can be influenced by superposed emission. This shows nicely in the middle plot in  Figure~\ref{absorption_correlation}, where RNO 1B is shifted with respect to the other objects. 

Finally, it is interesting to note that the objects showing evidence for ice processing in 
Figure~\ref{absorption15} tend to show higher optical depths in the ice features than the other
sources. This might suggest that the extinction for the latter objects (V346 Nor, V1735 Cyg) 
might be caused by ices somewhere in the line of sight to the source, 
rather than by material related to the young star. This might also explain why V1647 Ori shows
weak silicate emission, but ice absorption: the extinction is caused by cold foreground material,
reflected also in the observed high value for A$_{\rm V}$ (see caption Figure~\ref{emission}).
Ground-based observations find the spectrum to be flat in the 10\,$\mu$m regime \citep{abraham2006}.
However, the sensitivity of {\sc Spitzer} allows the detection of a weak feature 
resulting from an extinguished but intrinsically strong silicate emission band.


\subsubsection{Category 2 FUors: the silicate emission feature and its dust composition}
As presented in Figure~\ref{emission} and described in the related section, the emission profiles bear
evidence for dust grain processing. Even after the correction for apparent interstellar extinction, the 
shape of the silicate profile differs from that of typical ISM dust. Like for TTauri stars and the 
slightly more massive HAeBes, the origin of the emission feature in the 
spectra of the FUors is the heated surface layer of the accretion disk. The apparent grain
processing is believed to be only possible in circumstellar disks and not in he less dense circumstellar envelopes.
While the emission layer for TTauri stars and HAeBes is mainly heated be the central stellar object, for 
FUors the hot inner parts of the accretion disk itself can act as the main illuminating source
\citep[e.g.,][]{lachaume2004}. Due to the high accretion rates of FUors those inner regions are extremely 
hot and account for a significant fraction of the total flux even at optical wavelengths \citep{quanz2006}.
Furthermore, accretion disk models with a flared geometry were not only able to explain the SEDs 
of FU Ori and Bran 76 \citep{green2006}, but they could also reproduce interferometric observations in the 
NIR \citep{malbet2005} and MIR \citep{quanz2006}. V1057 Cyg and V1515 Cyg show more emission 
at longer infrared wavelengths than the previous objects, which can be accounted for assuming a remnant infalling 
envelope in addition to the accretion disks. This is supported by the results from K-band interferometry 
\citet{millan-gabet2006}, where envelopes are required to explain the low NIR visibilities. 

Concerning the dust composition it is interesting to note that, given the low peak-over-continuum ratio
illustrated in Figure~\ref{fluxratio}, grain growth must 
have already set in. This is supported by the results of our dust model fit.
Furthermore, the spectra do not show  
evidence for crystalline dust particles. The mass fraction of crystalline particles in 
the dust model computed for the spectrum of FU Ori was negligible (see, Table~\ref{dust_table}). 
Since strength and shape of the silicate feature of FU Ori is
comparable to those of Bran 76 and V1515 Cyg (Figure\ref{normalizedemission}), the dust composition
in all objects is similar. The spectrum of V1057 Ori is even broader and less pronounced than the other 
spectra, indicating even larger grains in the disk surface layer. The intrinsic feature of V1647 Ori 
is stronger than those of the other FUors, but no prominent signs of crystalline
silicates are present either.\footnote{The spectrum of Parsamian 21 does show a
prominent feature at 11.3\,$\mu$m but we attribute most of the related flux to 
PAH emission and not to crystalline silicates (see section~\ref{par21}).} 

As already mentioned in \citet{quanz2006}, there are several reasons 
for which stronger crystalline features could have been expected to be detected. 
The high disk accretion rates should ease the detection of crystalline particles in two ways:
(1) high accretion rates lead to high disk temperatures \citep{bell1997} which 
in turn should increase the amount of crystalline particles produced by annealing processes at T$\ge$800 K.
(2) an increase in the accretion rate should also increase 
the radial and vertical mixing in the disk \citep{gail2001}
transporting the crystalline particles farther out and to the disk surface, where they can 
be detected by means of MIR spectroscopy. However, neither in the innermost disk regions probed with 
MIR interferometry \citep{quanz2006} nor in the {\sc Spitzer} spectra presented here in 
Figure~\ref{emission} or at longer wavelengths in Figure~\ref{spitzer_long1} we see any striking 
evidence for crystallinity. This means, that either those grains do not exist in large amount in these
disks or they are somehow hidden.

The disks of FUor objects are different from those of TTauri or Herbig star
disks in the sense that in the radially innermost regions of FUor disks the luminosity
is accretion dominated (TTauri and Herbig star disks are irradiation
dominated everywhere). At larger radii, also FUor disks are irradiation
dominated, albeit the main heating source may not be the central star itself, but
rather the hot inner disk regions close to the star where accretion dominates \citep[e.g.,][]{lachaume2004}. 
Disk regions that are irradiation dominated
have a surface layer that is warmer than the underlying disk interior; such an
"inverted" temperature profile causes dust features to appear in emission.  In
the accretion dominated regions, the main heating source is the release of
gravitational energy in the disk midplane. Here, the disk interior is at least
as hot (and likely hotter) than the disk surface, causing dust features to be
effectively hidden, or even to appear in absorption\footnote{The gas absorption features
between 5 and 8\,$\mu$m described in 3.2.4 likely originate in this hot disk region.}. 
Thus, it is conceivable
that there are significant amounts of crystalline silicates present in the hot
inner disk regions of our targets, even if they do not show up prominently in
the spectra.

But what about crystalline silicates further from the central object, in
irradiation dominated regions?  Given the high accretion luminosity in our
objects it seems unavoidable that large amounts of crystalline silicates are
produced in the innermost disk regions. In addition to this, crystalline silicates may be produced \emph{in
  situ} at large radii in shocks \citep{harker2002} or electric
discharges \citep{pilipp1998,desch2000}. If the disk is
well mixed in the vertical direction, the crystalline silicates should be
present in the surface layer of the disk at large ($\gtrsim$\,few~AU) radii,
and show up in emission, which they evidently do \emph{not} (at least, to a
much lesser extent than in many TTauri and Herbig Ae/Be star disks, which are
more evolved). This inevitably leads to the conclusion that
crystalline silicates are \emph{not} abundant in the surface layer of
FUor disks at radii of more than a few AU from the central object,
\emph{contrary to expectations}. 

At this point, we have no proper explanation for this observation. In order to
animate the discussion, we will post the following idea, but stress that at
this point this is mere speculation. The crystalline silicates are not present
in the surface layer at larger radii, but may be present in the disk interior,
where they are expected to be formed by virtue of the processes mentioned
before. Thus, we need to explain why the crystalline silicates do not get
mixed in the vertical direction all the way to the disk surface. One
possibility is that the crystalline particles are somehow less well coupled to
the gas, possibly because they are more compact (or less "fluffy") than the
unprocessed dust particles. This may cause the selective settling of
crystalline dust.
A second possibility is the following. The accretion rate in FUor objects is
high, suggesting that the disk is still being supplied with fresh material
from the maternal envelope. This material, which is thought to contain
exclusively small, amorphous dust particles, will in part "rain" onto the disk
surface at larger radii. If the rate at which this material falls onto the
disk is higher than the rate at which it is mixed through the disk in the
vertical direction, the disk may be covered by a "blanket" of pristine
material, effectively hiding the more processed material present in the disk
interior. It is unclear whether this idea can be harmonized with the
observation that the dust we see at the disk surface has undergone significant
grain growth. This issue clearly calls for a much more detailed and
quantitative investigation, which is beyond the scope of the present work.

\subsection{Unifying the two categories of FUors: An evolutionary sequence}
The spectral properties of the sample presented here draw a rather
inhomogeneous picture of the group of FU Orionis objects. However, as most of them 
are convincingly classified as FUors, they do share some common observational and physical
properties. In the following, we present an idea, how the apparent differences can be 
explained within a unified paradigm of FU Orionis objects. 

Already 20 years ago \citet{herbig1977} and \citet{hartmannkenyon1985} suggested that each 
FUor might undergo several outbursts. More recent theoretical models show that
gravitational instabilities in the accretion disk, driven by continuing 
infall from a remnant envelope, can account for intense bursts of high accretion rates which 
are intersecting more prolonged, quiescent periods of low accretion \citep{vorobyovbasu2006,boley2006}.
After several of these outbursts have occurred within a time span of several 10$^5$ years, 
the envelope, which is the trigger of the disk instabilities, vanishes, and the object 
enters finally a state of permanently low accretion. If one relates this final phase to the
classical TTauri phase of low-mass YSOs, then FUors are younger than classical TTauri stars 
as already pointed out by \citet{weintraub1991} and \citet{sandell2001}.

\begin{figure}
\centering
\epsscale{1.}
    \plotone{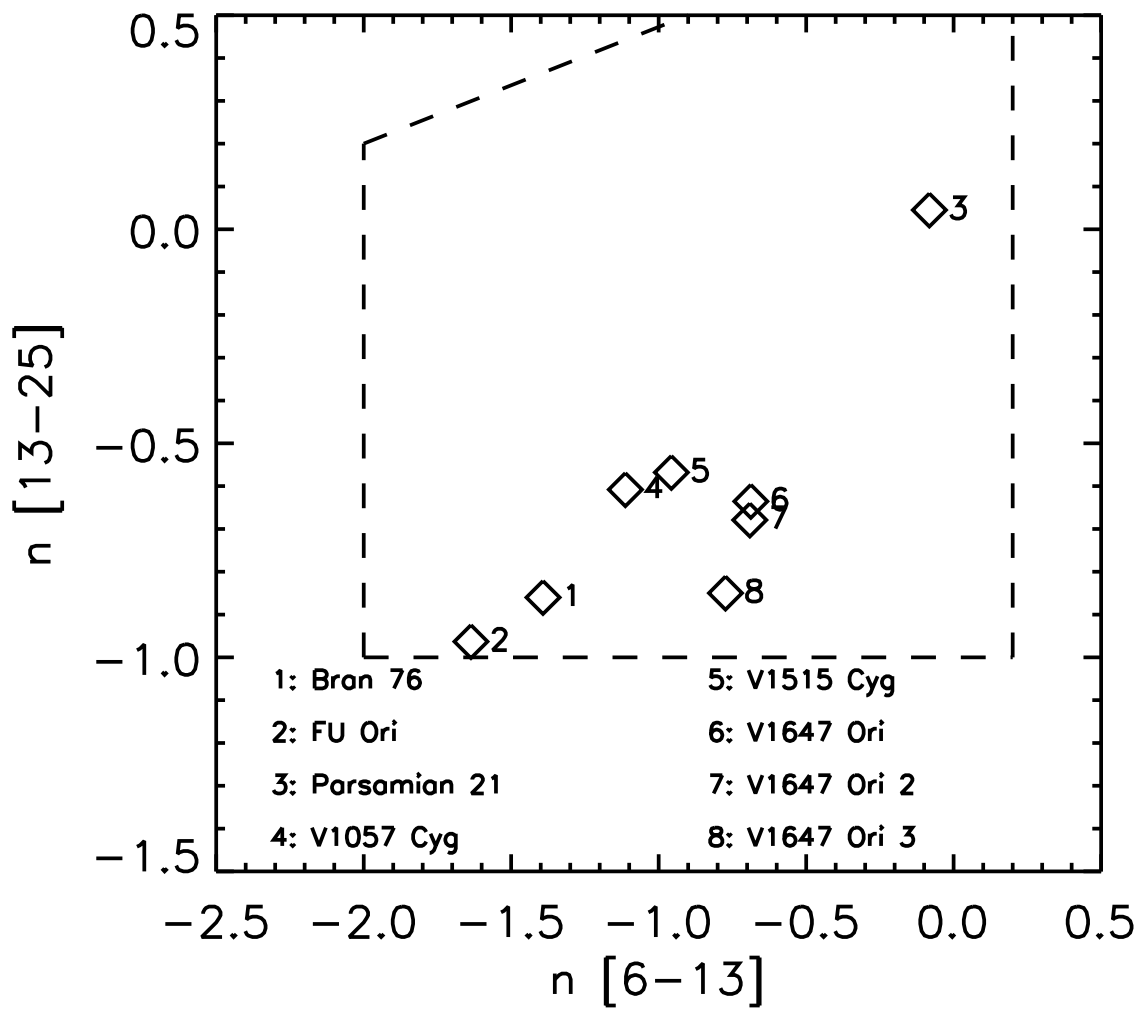}
   \caption{Spectral index $n$ of the silicate emission objects 
   evaluated between 13 and 25\,$\mu$m vs. the spectral index between 6 and 13\,$\mu$m.
   $n$ is computed as $n\equiv d\,{\rm log}(\lambda F_{\lambda})/d\,{\rm log}(\lambda)$. 
   All objects populate the same area in the plot as the Class II objects in Taurus 
   presented by \citet{furlan2006} (dashed area). Due to the steep rise in its SED, Parsamian 21 lies slightly 
   off from the other objects.}
 \label{spectral-indices}
\end{figure}

\begin{figure}
\centering
\epsscale{1.}
    \plottwo{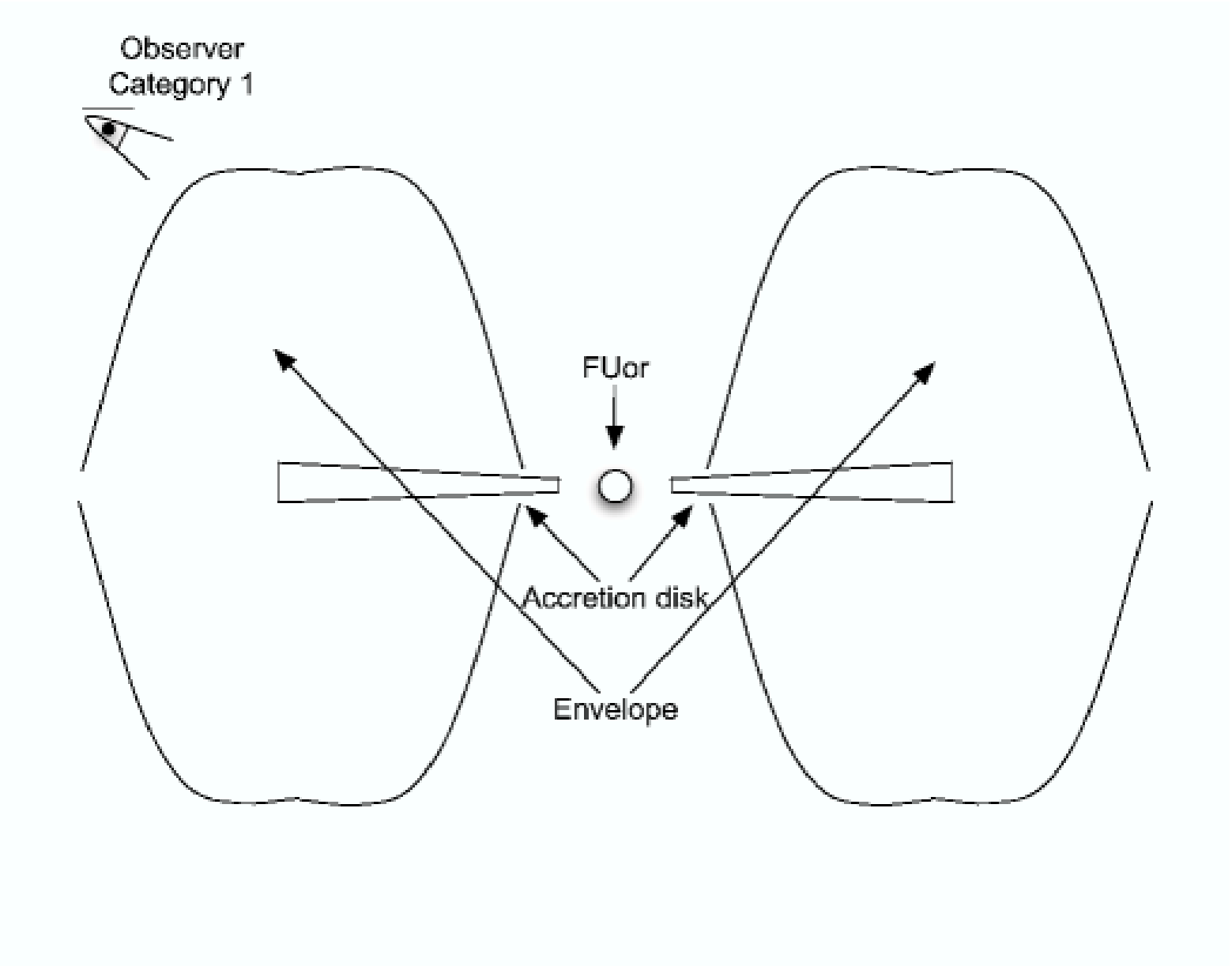}{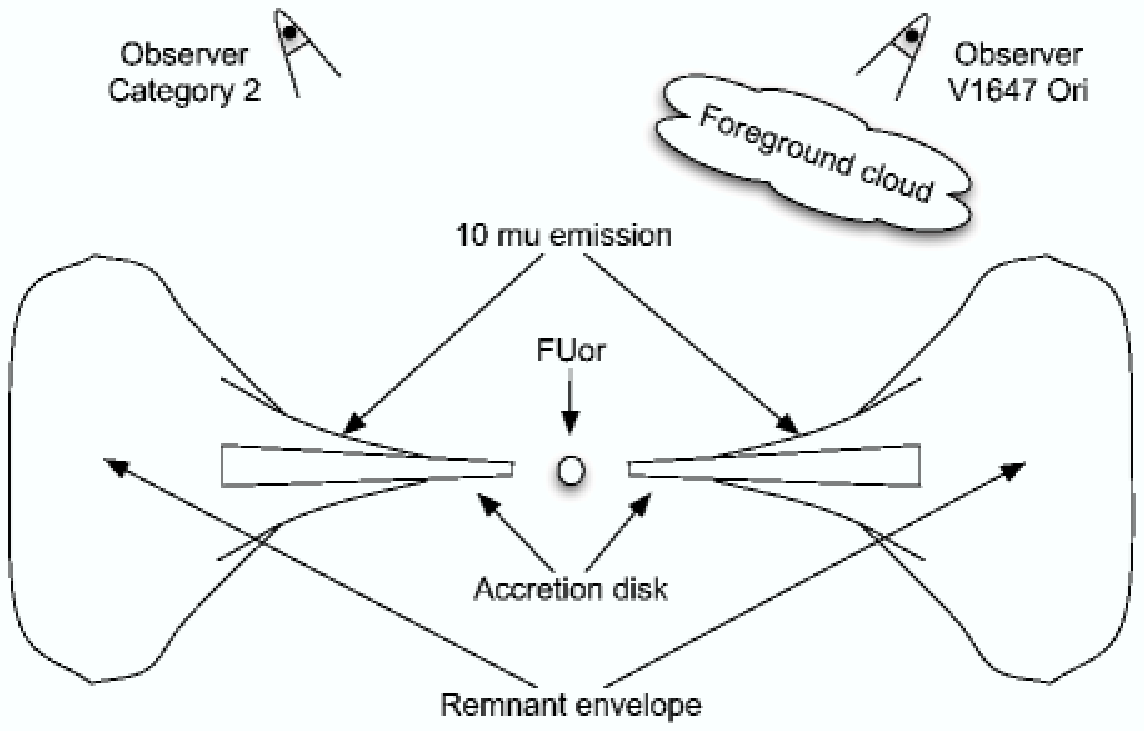}
   \caption{Simple graphical representation explaining the two 
   categories of FUors (not to scale). 
   Category 1 objects (left) show silicate absorption and are younger than than Category 2 
   objects (right) which show silicate emission. While Category 1 objects are still more deeply embedded in 
   their envelopes and appear to be Class I sources, Category 2 FUors are 
   similar to Class II sources.
   The spectrum of V1647 Ori can be explained assuming that extinction in the line of sight to this object
   creates the ice absorption features and reduces also the strength of the observed silicate emission.  
   }
 \label{categories}
\end{figure}

\citet{kessler2005} analyzed the 10\,$\mu$m silicate feature of 34 young stars and confirmed that 
the evolution of the feature of low-mass stars and the overall SEDs is similar to that of intermediate-mass stars:
embedded objects, showing a pristine silicate band in absorption, evolve into objects showing
a combination of silicate absorption and emission, and finally pure emission features appear, where 
dust processing leads to a broad range of shapes and strength. Since all these stages are 
represented in our sample, we believe that, indeed, the FUor-phase is the link between the 
embedded Class I objects and the more evolved Class II objects. In this context, the 
objects showing silicate absorption are younger and at the beginning of the period where 
subsequently numerous FUor outbursts will occur, while the objects showing pure emission features
are more evolved and possibly near the end of their FUor period. As illustrated in Figure~\ref{spectral-indices}
the spectral indices of the emission objects are indistinguishable from Class II objects in Taurus \citep{furlan2006}. In particular, the objects FU Ori and Bran 76 appear to be the most evolved
objects, because, as mentioned above, their SEDs do not require the presence of a large remnant envelope.
A nice example of an intermediate object is RNO 1B, where the dominating absorption feature is 
altered by the underlying silicate emission from the accretion disk. Figure~\ref{categories} illustrates with simple 
sketches the main features of the the two categories of FUors. It should be mentioned that 
theoretically a Class II FUor seen close to edge-on might be interpreted as a Class I FUor. As outlined
in 4.1.1 it is, however, statistically very unlikely that all of the Class I objects shown here are 
Class II FUors "in disguise".

One question arises, though. If FUors are younger than classical TTauri stars but do already show clear
evidence for dust coagulation, then why do some classical TTauri stars show silicate emission features 
consisting purely of pristine and unprocessed dust? One possible explanation is linked to the
high accretion rates of the FUors. It is believed that high disk accretion rates are related to a higher
degree of turbulence and thus mixing in the disk. As a result, larger particles are coupled to the turbulent
gas and mixed throughout the disk and to the disk surface, where they can be observed. Once the
accretion rate, and hence the amount of mixing, drops, larger particles tend to settle to the
disk midplane much faster than small particles, which remain in the disk surface layer. Thus, although 
large particles are present in the disk of a TTauri star, they might not reach the disk surface, and only
the small grains produce a silicate emission feature. Again, as for the issue concerning the
lack of crystalline grains, a more detailed theoretical modeling is certainly
required to see to what extend this qualitative explanation is valid.

\subsection{Parsamian 21 - intermediate mass FUor or Post-AGB star?}\label{par21_discussion}
The strong PAH emission of Parsamian 21 is unique among the sample of FUors presented here, and the
analysis rise serious doubts whether Parsamian 21 is indeed a member of this group. The initial classification 
was made by \citet{staudeneckel1992} based on optical spectroscopy. These authors derived a spectral 
type of F5Iab, found a prominent P-Cygni profile in H$\alpha$ and observed shock-induced emission in [O I],
[N II] and [S II]. They also mention the detection of Li in their spectra, which, if present, does not
seem to be significantly above the noise level in the data. 
New observations with higher spectral resolution and higher signal-to-noise would certainly be 
eligible. Although some of these features are commonly observed in FUors, they are not unique to this group 
of objects, as also HAeBes show, for instance, P-Cygni profiles in H$\alpha$. 
In addition, two of the main properties of FUors, an outburst in optical light and CO bandhead 
absorption profiles in the NIR \citep[e.g.,][]{hartmann1996}, have not yet been observed for Parsamian 21. 
Looking at the immediate surroundings of Parsamian 21 in optical data from the Digitized Sky Survey\footnote{http://archive.eso.org/dss/dss} it is also
striking that no dark cloud complex is linked to Parsamian 21 and hence no connection to a larger
star-forming region is present.

To excite the PAH emission, sufficiently strong UV radiation is required. However, most 
FUors have a later spectral type than Parsamian 21 and lack, unlike TTauri stars, in general UV continuum 
excess emission \citep{hartmann1996}.
As mentioned above, the positions of the PAH emission bands are untypical
for young stars and so far only 7 objects are known to show characteristics of PAH class C objects \citep{sloan2007} as defined by \citet{peeters2002}. Among these 7 objects only two presumably young stars (SU Aur and HD 135344) 
show the peculiar PAH band close to 8.2\,$\mu$m. All other PAH class C objects are
post-AGB stars or red giants \citep{sloan2007,peeters2002}. 

In addition, one of the Post-AGB stars from 
\citet{peeters2002} and also HD 56126, a Post-AGB star with PAH and silicate emission \citep{hony2003}, have
the same spectral type as Parsamian 21. However, we should mention 
that \citet{dibai1969} and \citet{the1994} give a 
spectral type of A5Ve$\alpha$ for Parsamian 21. This is too early for a typical 
FU Orionis object, but could also explain the PAH emission.
Finally, the bipolar emission knots seen in H$\alpha$ and [N II] in the optical spectrum of Parsamian 21 \citep{staudeneckel1992}, are also found in the bipolar outflows of evolved stars and "pre-planetary" nebulae
\citep[e.g., in the "Butterfly Nebula", ][]{solf2000}. They are thus not only seen in young objects.

Based on these findings, we believe that the FUor status of Parsamian 21 is at least very questionable.
Either this object represents an intermediate mass FUor object, suggesting that also stars in this 
mass regime undergo phases of enhanced accretion, or, and this appears even more likely, Parsamian 21
is not even a young object but an evolved star, sharing typical properties with Post-AGB stars.

\section{Conclusions and future prospects}
Our conclusions can be summarized as follows:
\begin{itemize}
\item{We presented the first coherent space based spectroscopic MIR study 
of 14 FUors observed with the {\sc Spitzer Space Telescope} or the
{\sc Infrared Space Observatory}. The sample includes roughly two thirds of the known FUors or FUor candidates.}
\item{Based on the appearance of the 10\,$\mu$m silicate feature, we divided the sample into two categories: Category 1 objects
show the silicate feature in absorption, and the spectra show additional absorption bands at 6.0, 6.8 and 15.2\,$\mu$m due to ices. 
Category 2 objects show silicate 
emission and (most of them also) indications for water vapor absorption at shorter wavelengths. Only one Category 2 object (V1647 Ori) 
shows ice absorption bands which we
explain by foreground extinction and an intrinsically strong 10\,$\mu$m emission feature.}
\item{The silicate absorption is best explained with dusty and icy envelopes surrounding the Category 1 objects. Statistical 
reasons argue against all objects having accretion disks seen edge-on. The silicate emission of the Category 2 FUors arises from 
the surface layer of the surrounding accretion disks.}
\item{The shape of the silicate band of the Category 1 objects is in agreement with typical dust compositions of the ISM. For one object
(RNO 1B)  the shape of the feature and the decrease in optical depth longward of 10\,$\mu$m can be explained with a superposed emission feature.}
\item{Optical depths for the observed silicate and ice absorption bands were derived. We find an apparent correlation among the optical depths of the 
ices indicating similar environmental conditions for the objects. For the silicate feature, no correlation is expected.}
\item{Using different extinction curves we computed the optical extinction 
toward the objects based on the depth of the 10\,$\mu$m silicate features. The results are in agreement with values found in the literature, given the uncertainties in the dust models and for the optical 
extinction values in the references.
For RNO 1B the derived value for A$_{\rm V}$ is smaller than expected, indicating that on-top silicate emission might influence the optical depth at 10\,$\mu$m.}
\item{The emission profiles of the Category 2 objects show clear evidence for grain growth. Fitting a dust model to the spectrum of FU Ori reveals that, indeed,
larger grains than typically observed in the ISM are required to explain the shape. As FUors are presumably younger than TTauri stars (see below), this 
indicates that grain growth sets in very early during disk evolution.}
\item{Despite the high accretion rates of the FUor accretion disks and the resulting higher disk temperatures and mixing rates, we find hardly any evidence
for crystalline grains. So far we lack a clear explanation for this observational results and leave it to further investigations.}
\item{The two categories of FUors can be explained within a single paradigm, where Category 1 objects are younger and similar to Class I objects, while
Category 2 FUors are more evolved and show already properties of Class II sources. This explanation is in agreement with theoretical models which 
expect FUors to undergo several outbursts before they enter the more quiescent classical TTauri phase. Thus, the FUor-phase might indeed be the link between
Class I and Class II objects and common to most young low-mass stars.}
\item{For OO Ser and V1647 Ori the multi-epoch data allows an analysis of the post-outburst fading of the objects. Both objects fainted significantly 
over timescales of a few months, suggesting the outbursts might be intermediate between the long-lived FUor-eruptions and the short-lived EXor-type events.}
\item{Only one object (Parsamian 21) shows PAH emission similar to that often observed in Post-AGB stars. 
We find that most other observational data for Parsamian 21
can also be explained with the object being an evolved star. In consequence, the FUor-status for this object is questioned.  }
\end{itemize}

Based on these findings, future investigations might include the following points:
\begin{itemize}
\item{A complete MIR spectroscopic census of all known FUors would complement the data presented here and might
help to derive conclusions concerning the duration of the FUor phase using statistical arguments.}
\item{Multidimensional radiative transfer models have not yet been applied to FUor accretion disk, 
but they are required to 
derive a coherent picture of the disk structure, including the emission layer.}
\item{Models for the dust evolution in accretion disks could try to explain quantitatively the observed large grain population and the
apparent lack of crystalline silicates.}
\item{Finally, if our conclusions are correct, then some of the known low-mass Class I objects might be FU Ori objects hidden in a 
quiescent phase between two consecutive outbursts. The observational properties of these objects might be revisited to search for any
indication of FUor properties.}
\end{itemize}







\acknowledgments
S.~P.~Q.~kindly acknowledges support from the German
\emph{Friedrich-Ebert-Stiftung}. We are grateful to Henrik Beuther and Kees Dullemond for 
interesting and insightful discussions and thank the referee for a detailed report helping to
improve the style and content of this paper.  
The version of the ISO data presented in this paper correspond to Highly Processed Data Product (HPDP) sets  available for public use in the ISO Data Archive. OSIA is a joint development of the SWS consortium. Contributing institutes are SRON, MPE, KUL and the ESA Astrophysics Division. This research has made use of the SIMBAD database,
operated at CDS, Strasbourg, France. 



{\it Facilities:} \facility{Spitzer},\facility{ISO}

\clearpage

\begin{deluxetable}{lcccccl}
\rotate
\tablecaption{Journal of {\sc ISO-SWS} observations. The coordinates denote the pointing
position of the telescope as saved in the header of the data files. The speed of the observations, the time spent on the target and possible pointing offsets in the telescope's y and z axis are given as well. The offsets were
corrected for during the data reduction applying the measured beam profiles along the different axes. 
\label{isojournal}}           
\tablewidth{0pt}
\tablehead{
\colhead{Object} & \colhead{RA (J2000)} & \colhead{DEC (J2000)} & 
\colhead{AOT / Speed} & \colhead{Time on} & \colhead{Offset [$''$]} & \colhead{Date} \\
	 &  &     &  &	\colhead{target [sec]}  &  \colhead{y / z}  & 
}
\startdata
\object{OO Ser}  &  18h29m49.05s & +01d16$'$19.2$''$ & SWS01 / 1 &  1062 &  - / - &1996-04-14 \\  
               &  18h29m49.08s   & +01d16$'$19.8$''$ & SWS01 / 1 &  1140 &  4 / 3 &1996-10-24 \\
  	     &  18h29m49.05s & +01d16$'$19.2$''$ & SWS01 / 2 &  1140 &  - / - & 1997-03-08 \\
  	    &  18h29m49.05s & +01d16$'$19.2$''$ & SWS01 / 3 &  3454 &  - / - & 1997-04-12 \\
  	    &  18h29m49.09s & +01d16$'$19.8$''$ & SWS01 / 3 &  3454 &  - / - & 1997-09-22 \\
RNO 1B  & 00h36m46.24s & +63d28$'$54.3$''$	& SWS01 / 2 & 	1912 &  - / - &  1996-08-27\\
V346 Nor &  16h32m32.05s & -44d55$'$28.9$''$ & SWS01 / 2 & 1912 &  - / - &  1996-08-31\\
\object{Z CMa} &	 07h03m43.17s &  -11d33$'$06.6$''$ & SWS01 / 2 &  3454 &   3 / 3 &  1997-11-07\\
V1735 Cyg  & 21h47m20.60s & +47d32$'$04.9$''$ & SWS01 / 2 & 1912 &  4 / 4 & 1996-08-06\\
Reipurth 50 & 05h40m17.89s &  -07d27$'$29.3$''$	& SWS01 / 3 &	3454 &  5 / 5 & 1997-10-13\\
L1551 IRS 5 & 04h31m34.06s & +18d08$'$04.8$''$	& SWS01	/ 4 & 	6538 & - / -  & 1997-09-06\\
 
\enddata
\end{deluxetable}



\begin{deluxetable}{lcclll}
\rotate
\tablecaption{Journal of {\sc Spitzer-IRS}  observations. The coordinates denote the average slit 
position of the low resolution spectrograph as computed by the on board software. 
The AOR of the observations, the time spent on the target for the different modules and the observation date are given as well. 
\label{spitzerjournal}}           
\tablewidth{0pt}
\tablehead{
\colhead{Object} & \colhead{RA (J2000)} & \colhead{DEC (J2000)} & 
\colhead{AOR} & \colhead{Integration time $[$sec$]$} & 
\colhead{Date} 
}
\startdata
\object{Bran 76} (BBW 76) & 07h50m35.52s   & -33d06$'$24.12$''$ & 3571200 &  12 (SL, LL)   & 2004-04-14 \\  
\object{FU Ori}  &  05h45m22.39s   & +09d04$'$12.5$''$ & 3569920 &  12 (SL, SH, LH)   & 2004-03-04 \\
\object{L1551 IRS 5}  &  04h31m34.08s & +18d08$'$04.92$''$ & 3531776 &  6 (SL, SH, LH) &  2004-03-04 \\
\object{Parsamian 21} (HBC 687) &  19h29m00.72s & +09d38$'$47.11$''$ &  5039872 &  36 (SL) &     2004-04-18 \\
             &              &                   &          & 48 (SH, LH)   &        \\
\object{RNO 1B}   &  00h36m46.34s &  +63d28$'$53.76$''$    & 6586624  &  36 (SL) &    2004-01-07\\
             &              &                   &          &  60 (SH, LH)   &         \\
\object{RNO 1C}   &  00h36m46.89s &  +63d28$'$58.44$''$    & 6586624  &  36 (SL) &     2004-01-07 \\
             &              &                   &          &  60 (SH, LH)   &        \\
\object{V1057 Cyg} & 20h58m53.76s & +44d15$'$28.44$''$     & 3570176  &  12 (SL, SH, LH)&   2003-12-15\\
\object{V1515 Cyg} & 20h23m48.00s & +42d12$'$25.56$''$     & 3570432  &  12 (SL, SH, LH)&   2004-05-11 \\

\object{V1647 Ori}\tablenotemark{a} & 05h46m13.13s & -00d06$'$05.21$''$     & 12261120 &  12 (SL) &     2004-10-20 \\
             &              &                   &          &  24 (SH, LH)   &          \\
          & 05h46m13.15s & -00d06$'$04.41$''$      & 11569920 & 48 (SL) &     2005-03-11 \\
             &              &                   &          &  484 (SH), 240 (LH)   &       \\
          & 05h46m13.14s & -00d06$'$04.69$''$       & 12644096 &  12 (SL, LL) &     2005-03-24 \\
\object{V1735 Cyg} & 21h47m20.6s &  +47d32$'$00.7$''$	& 3570944  &  12 (SL, SH, LH) &  2003-12-17 \\
\object{V346 Nor}  & 16h32m32.1s &  -44d55$'$28.6$''$	& 3570688  &  12 (SL, SH, LH) & 2004-02-27 \\\hline
\object{HL Tau} & 04h31m38.4s &  +18d13$'$57.9$''$	& 3531776  &  6 (SL, SH, LH) & 2004-03-04 \\
\object{XZ Tau} AB& 04h31m40.1s &  +18d13$'$57.4$''$	& 3531776  &  6 (SL, SH, LH) & 2004-03-04 \\
\enddata
\tablenotetext{a}{The object was observed at three different epochs.}
\end{deluxetable}

\thispagestyle{empty}
\begin{deluxetable}{llccccccc}
\rotate
\tablecaption{Overview of prominent spectroscopic features (mostly ices and dust) 
seen in the spectra presented in Figures~\ref{iso_short1},~\ref{iso_short2},~\ref{spitzer_short1},~\ref{spitzer_short2},~\ref{spitzer_long1},~\ref{spitzer_long2}. "abs" denotes an absorption feature, "em" an emission feature. For objects where
ISO and {\sc Spitzer} data are available, the {\sc Spitzer} data is of higher quality and thus more reliable.  
\label{features}}           
\tablewidth{0pt}
\tablehead{
\colhead{Object} & 
\colhead{Instrument} &
\colhead{H$_2$O} (ice) & 
\colhead{H$_2$CO(?)/HCOOH(?)/} &
\colhead{H$_2$O (gas)} &
\colhead{CH$_3$OH(?)/} &
\colhead{silicates} &
\colhead{CO$_2$ (ice)} &  
\colhead{PAH\tablenotemark{a}} \\
 & 
 &
 & 
\colhead{H$_2$O (ice)} &
 &
\colhead{NH$_4^+$(?)} &
 &
 &  
 \\
 &   &
\colhead{3.08$\mu$m} & 
\colhead{5.85 / 6.0\,$\mu$m} & 
\colhead{5.8 / 6.8$\mu$m}&
\colhead{6.85$\mu$m} &
\colhead{10.0$\mu$m} &
\colhead{15.2$\mu$m} & 
}
\startdata
OO Ser       &  ISO-SWS       &   - & -   & -  &  -   & abs & abs & - \\
V346 Nor     &  ISO-SWS       & abs & abs & -  &  -   & abs & -   & - \\
             &  {\sc Spitzer} IRS   &   - & abs & -  &  abs & abs & abs & - \\
Z CMa         &  ISO-SWS       &   - & -   & -  &  -   & abs & -   & - \\
Reipurth 50  & ISO-SWS        & abs & abs & -  &  abs & abs & -   & - \\
L1551 IRS 5  & ISO-SWS        & abs & abs & -  &  abs & abs & abs & - \\
             & {\sc Spitzer} IRS    &   - & abs & -  &  abs & abs & abs & - \\
RNO 1B       &  ISO-SWS       & abs & -   & -  &  -   & flat & - & - \\
RNO 1B       &  {\sc Spitzer} IRS   &   - & abs & -  &  abs & abs & abs & - \\
RNO 1C       &  {\sc Spitzer} IRS   &   - & abs & -  &  abs & abs & abs & - \\
V1735 Cyg    &  ISO-SWS       & abs & abs?& -  &  -   & flat & - & - \\
             &  {\sc Spitzer} IRS   &   - & -   & -  &  -   & abs  & abs? & - \\
Parsamian 21 & {\sc Spitzer} IRS    &   - & -   & -  &  -   & em  & -  & em \\  
Bran 76 (BBW 76)       & {\sc Spitzer} IRS    &   - & -   & abs&  -   & em  & -  & -  \\
V1057 Cyg    & {\sc Spitzer} IRS    &   - & -   & abs&  -   & em  & abs (gas) & - \\
V1515 Cyg    & {\sc Spitzer} IRS    &   - & -   & abs&  -   & em  & -   & -  \\
V1647 Ori    & {\sc Spitzer} IRS    &   - & abs & -  &  abs & em  & abs & - \\
FU Ori       & {\sc Spitzer} IRS    &   - & -   & abs&  -   & em  & -   & -  \\\hline
V883 Ori\tablenotemark{b}     & ESO Timmi2     &   - & -   & -  &   -  & abs & -   & - \\\hline
XZ Tau AB    & {\sc Spitzer} IRS    &   - & -   & -  &  -   & em  & -   & - \\
HL Tau       & {\sc Spitzer} IRS    &   - & abs & -  &  abs & abs   & abs & - \\
\enddata 
\tablenotetext{a}{Any significant emission at 6.2 or 7.7\,$\mu$m (C-H modes), 8.6, 11.3 or 12.7\,$\mu$m (C-C modes)
or at 8.2\,$\mu$m (origin not clear yet).}
\tablenotetext{b}{Observations from \citet{schuetz2005}.}
\end{deluxetable}


\begin{deluxetable}{lcccc}
\tablecaption{Temperatures and power law indices for the different components of the 
analytical dust model for FU Ori shown in Figure~\ref{dust_model}.  
\label{temperature_table}}           
\tablewidth{0pt}
\tablehead{
\colhead{Component} & \colhead{T [K]}& \colhead{Exponent $p$}
}
\startdata
Inner disk rim  & 2246 & -0.84 \\
Disk surface & 1128 & -0.49  \\
Disk midplane & 865 & -0.11 \\
\enddata
\end{deluxetable}


\begin{deluxetable}{lcccc}
\tablecaption{Mass fractions of dust species with different grain sizes 
derived from the dust model fit for FU Ori shown in Figure~\ref{dust_model}.  
\label{dust_table}}           
\tablewidth{0pt}
\tablehead{
\colhead{Species} & \colhead{0.1\,$\mu$m}&\colhead{1.5\,$\mu$m}& \colhead{6.0\,$\mu$m}& \colhead{Total}
}
\startdata
Amorphous olivine & 0.16 & 0.20 & 0.21 &  0.57 \\
Amorphous pyroxene & 0.00 & 0.00 & 0.43 & 0.43 \\
Crystalline forsterite & $<$0.01 & 0.00 & 0.00 & $<$0.01 \\
Crystalline enstatite & 0.00 & 0.00 & 0.00 & 0.00\\
Silica & 0.00 & 0.00 & 0.00 & 0.00\\
Amorphous carbon & 0.00 & 0.00 & 0.00 & 0.00\\\hline\\
Total & 0.16 & 0.20 & 0.64 & \\
\enddata
\end{deluxetable}


\begin{deluxetable}{lccccc}
\tablecaption{Optical depths and extinction values as derived from the spectra shown in Figure~\ref{absorption}.
The last column list reference values for $A_{\rm V}$ from the literature.
\label{extinctiontable}}           
\tablewidth{0pt}
\tablehead{
\colhead{Object} & \colhead{$\tau_{\rm silicate}$} & \colhead{$\lambda_{\rm silicate}$ [$\mu$m]} & 
\colhead{$A_{\rm silicate}$ [mag]} & \colhead{$A_{\rm V}$ [mag]} & \colhead{$A_{\rm V}^{literature}$ [mag]}  
}
\startdata
     L1551 IRS 5 &  1.33$\pm$0.03  & 9.44 &  1.44$\pm$0.03 &   21.83$\pm$   4.70 &
      19-20\tablenotemark{a}, 30\tablenotemark{b}, $\gtrsim$150\tablenotemark{c} \\
      
       V346 Nor  &  0.32$\pm$0.01  & 9.68 &  0.35$\pm$0.01 &    4.98$\pm$   0.86 &  
       6.2\tablenotemark{d}\\
      V1735 Cyg  &  0.29$\pm$0.03  & 9.52 &  0.31$\pm$0.03 &    4.60$\pm$   0.91 &  
       10.0\tablenotemark{e,}\tablenotemark{f}\\
         RNO 1B  &  0.24$\pm$0.01  & 9.30 &  0.27$\pm$0.01 &    4.15$\pm$   1.03 &
	 9.2\tablenotemark{g}  \\
         RNO 1C  &  0.63$\pm$0.01  & 9.58 &  0.69$\pm$0.01 &   10.11$\pm$   1.87 &  
	 $\gtrsim$12\tablenotemark{g}  \\
         HL Tau  &  0.47$\pm$0.01  & 9.61 &  0.51$\pm$0.01 &    7.38$\pm$   1.32 &  
	 $\approx$24.0\tablenotemark{h,}\tablenotemark{i}, $\approx$38\tablenotemark{j}\\
          Z CMa  &  0.31$\pm$0.06  & 9.48 &  0.34$\pm$0.20 &    5.05$\pm$   1.04 &  
	 2.8$\pm$0.1\tablenotemark{k}  \\
        Reipurth 50 &  1.68$\pm$0.34  & 9.68 &  1.82$\pm$0.20 &   26.18$\pm$   4.34 &  
     $\approx$50\tablenotemark{l}  \\
\enddata
\tablenotetext{a}{\citet{snell1985}}
\tablenotetext{b}{\citet{smith1987}}
\tablenotetext{c}{\citet{campbell1988}}
\tablenotetext{d}{\citet{grahamfrogel1985}, assuming E(B-V)=2.0}
\tablenotetext{e}{\citet{sato1992}}
\tablenotetext{f}{\citet{levreault1988}}
\tablenotetext{g}{\citet{staudeneckel1991}, based on optical data and assuming spectral type.}
\tablenotetext{h}{\citet{bergin2005}}
\tablenotetext{i}{\citet{close1997}}
\tablenotetext{j}{\citet{menshchikov1999}, derived from radiative transfer model.}
\tablenotetext{k}{\citet{cohenkuhi1979}, based on optical data and assuming spectral type.}
\tablenotetext{l}{\citet{casali1991}, based on NIR colors, assumed intrinsic color temperatures and
strength of additional ice absorption features.}
\end{deluxetable}


\begin{deluxetable}{lccc}
\tablecaption{Optical depths of the main ice features at 6.0, 6.85 and 15.2\,$\mu$m. 
\label{iceextinctiontable}}           
\tablewidth{0pt}
\tablehead{
\colhead{Object} & \colhead{$\tau_{6.0\,\mu{\rm m}}$}&\colhead{$\tau_{6.85\,\mu{\rm m}}$}& \colhead{$\tau_{15.2\,\mu{\rm m}}$}
}
\startdata
     L1551 IRS 5 &  0.51$\pm$0.01	&0.47$\pm$0.01	&	0.54$\pm$0.05	\\
       V346 Nor  &  0.05$\pm$0.01	&0.06$\pm$0.01	&	0.12$\pm$0.02	\\
      V1735 Cyg  &  0.05$\pm$0.02	&0.09$\pm$0.01	&	$\approx$0.3 	\\
         RNO 1B  &  0.20$\pm$0.01	&0.18$\pm$0.01	&	0.32$\pm$0.04	\\
         RNO 1C  &  0.35$\pm$0.01	&0.26$\pm$0.01	&	0.41$\pm$0.02	\\
         HL Tau  &  0.10$\pm$0.01	&0.07$\pm$0.01	&	0.13$\pm$0.06	\\ 
      V1647 Ori  &  0.07$\pm$0.01	&0.06$\pm$0.01	&	0.20$\pm$0.02	\\
      V1647 Ori 2&  0.10$\pm$0.01	&0.08$\pm$0.01	&	0.19$\pm$0.02 	\\
      V1647 Ori 3&  0.09$\pm$0.01	&0.07$\pm$0.01	&	-\tablenotemark{a}  \\
\enddata
\tablenotetext{a}{only low-resolution data}
\end{deluxetable}


\begin{deluxetable}{lcccc}
\tablecaption{\emph{ Upper half:} Flux densities at 20 and 30\,$\mu$m and the slope of the SED between 15 and 30\,$\mu$m 
for OO Ser observed at 5 different epochs. \emph{ Lower half:} Flux densities at 8, 20 and 30\,$\mu$m for V1647 Ori at three 
different epochs. 
\label{ooserflux}}           
\tablewidth{0pt}
\tablehead{
\colhead{Date} &  \colhead{Flux density} & \colhead{Flux density} &\colhead{Flux density} & \colhead{Slope [Jy/$\lambda$]}\\
\colhead{} &  \colhead{at 8\,$\mu$m [Jy]} &\colhead{at 20\,$\mu$m [Jy]} &\colhead{at 30\,$\mu$m [Jy]} & \colhead{}
}
\startdata
\multicolumn{5}{c}{OO Ser}\\\hline\\
     1996-04-14 &  &16.5$\pm$1.8		&43.0$\pm$2.3	&	2.65$\pm$0.04 \\
     1996-10-24 &  &15.2$\pm$2.1		&41.2$\pm$2.6	&	2.61$\pm$0.05 \\
     1997-03-08 &  &12.8$\pm$1.2		&32.5$\pm$1.5	&	1.98$\pm$0.03 \\
     1997-04-12 &  &11.9$\pm$1.5		&31.1$\pm$1.8	&	1.92$\pm$0.04 \\
     1997-09-22 &  &8.8$\pm$1.2		        &23.9$\pm$1.4	&	1.52$\pm$0.03 \\\hline\\
\multicolumn{5}{c}{V1647 Ori}\\\hline\\
     2004-10-20 & 3.95$\pm$0.02 & 11.56$\pm$0.16 & 16.52$\pm$0.51 & \\
     2005-03-11 & 2.49$\pm$0.01 & 7.21$\pm$0.20 & 10.35$\pm$0.24 & \\
     2005-03-24 & 3.04$\pm$0.01 & 7.82$\pm$0.17\tablenotemark{a} & 10.62$\pm$0.05\tablenotemark{a} & \\
\enddata
\tablenotetext{a}{\phantom{}low-resolution data}
\end{deluxetable}




\clearpage

\end{document}